\documentclass[preprint,nofootinbib,
                 showpacs,preprintnumbers,amsmath,amssymb]{revtex4}
 
\usepackage{latexsym} 
 
\begin{document}     
\title{\bf Nonlocal braneworld action: an alternative to Kaluza-Klein 
description} 
\author{Andrei O.\ Barvinsky} 
\email{barvin@td.lpi.ru} 
\affiliation{Theory Department, Lebedev Physics Institute, 
Leninsky Pr. 53, 
Moscow 117924, Russia.} 
\author{Alexander Yu.\ Kamenshchik} 
\email{sasha.kamenshchik@centrovolta.it} 
\affiliation{ 
L. D. Landau Institute for Theoretical Physics of Russian 
Academy of Sciences, Kosygina str. 2, Moscow 117334, Russia,\\ 
and Landau Network --- Centro Volta, Villa Olmo, via Cantoni 1, 22100 Como, 
Italy.} 
\author{Andreas Rathke} 
\email{andreas.rathke@physik.uni-freiburg.de} 
\affiliation{Fakult\"at f\"ur Physik, Universit\"at Freiburg, 
  Hermann-Herder-Str. 3, 79104 Freiburg, Germany,\\ 
and Institut f\"ur Theoretische Physik, Universit\"at zu K\"oln,  
Z\"ulpicher Str. 77, 50937 K\"oln, Germany.} 
\author{Claus Kiefer} 
\email{kiefer@thp.uni-koeln.de} 
\affiliation{ 
Institut f\"ur Theoretische Physik, Universit\"at zu K\"oln,  
Z\"ulpicher Str. 77, 50937 K\"oln, Germany.} 
%\date{} 
\begin{abstract} 
We construct the nonlocal braneworld action in the  
two-brane Randall-Sundrum model in a holographic setup alternative to  
Kaluza-Klein description: the action is written as a functional  
of the two metric and radion fields on the branes. This action  
effectively describes the dynamics of the gravitational 
field both on the branes and in the bulk in terms of the brane geometries 
directly accessible for observations. Its nonlocal form factors  
incorporate the cumulative effect of the bulk Kaluza-Klein modes.  
We also consider the reduced version of this action obtained by  
integrating out the fields on the negative-tension brane invisible 
from the viewpoint of the Planckian brane 
observer. This effective action features a nontrivial transition  
(AdS flow) between the local and nonlocal phases of the theory  
associated with the limits of small and large interbrane separation.  
Our results confirm a recently proposed braneworld scenario 
with diverging (repulsive) branes and suggest possible new  
implications of this phase transition in brane cosmology.

\end{abstract} 
 
\maketitle 
\section{Introduction\label{intro}} 
Recent developments in string theory 
\cite{string} and the attempts to resolve the hierarchy problem 
\cite{hierarchy} suggest that the observable world can be a brane 
embedded in a higher-dimensional spacetime with a certain number 
of noncompact dimensions. Moreover, string-inspired field theories 
imply the existence of several branes interacting and propagating 
in the multi-dimensional bulk. Their dynamics manifests itself for 
the observer as an effective four-dimensional theory that, in the 
cosmological context \cite{BDL,ShiMaSa}, should explain the origin 
of structure in the Universe by means of an inflationary or some 
other scenario \cite{Ekpyr,Pyr}, explain its particle 
phenomenology, and shed light on problems such as a possibly 
observable cosmological acceleration \cite{accel}. 
 
The efficient way of description for the braneworld scenario is 
the method of effective action. Generally, this notion is very 
ambiguous, because its precise meaning ranges from the generating 
functional of one-particle irreducible diagrams in local field 
theory to the low-energy effective action of target spacetime 
fields in string theory. What is, apparently, in common for all 
these definitions is that the effective action manifestly describes 
the dynamics of the observable variables and, simultaneously, 
incorporates in implicit form the effect of invisible degrees of 
freedom that are integrated (or traced) out. This type of 
description becomes actually indispensable in string-theoretical, 
Kaluza-Klein and braneworld contexts when the observable variables 
turn out to be very different from the fundamental degrees of 
freedom whose dynamics underlies the effective ``visible'' 
dynamics. 
 
This situation is characteristic of the old Kaluza-Klein and new (braneworld) 
pictures, because the ``visible'' fields $\phi(x)$ are four-dimensional in 
contrast to the fundamental fields $\Phi(x,y)$ in the multi-dimensional 
spacetime depending on the visible (four-dimensional) coordinates $x$ and the 
coordinates of extra dimensions $y$. This is also the case in the widely 
celebrated by string physicists duality relations of the  
Anti-de\,Sitter/conformal-field-theory correspondence 
(AdS/CFT-correspondence) between the bulk and boundary theories  
\cite{AdS/CFT}. However, the four-dimensional fields $\phi(x)$  
originate from $\Phi(x,y)$ in these two cases by two different  
procedures and their effective actions essentially differ. 
 
In the Kaluza-Klein formalism $\phi(x)=\{\phi_n(x)\}$ arises as an  
infinite 
tower of Kaluza-Klein modes --- the coefficients of the expansion of 
$\Phi(x,y)$ in a certain complete set of harmonics on the $y$-space. 
Correspondingly, its effective action is just the original action of the 
fundamental field $S[\,\Phi(x,y)\,]$ rewritten in terms of $\{\phi_n(x)\}$. 
This type of action was built for the two-brane Randall-Sundrum scenario 
(cf.~\cite{RS}) in \cite{KubVol}, and a good review of its particle 
phenomenology can be found in \cite{Kub}.  This  action is, however,  
very often not 
helpful in the braneworld context, because it does not convey a 
number of its important features like non-compactness of extra 
dimensions \cite{RSloc,Rub}, recovery of the four-dimensional 
Einstein gravity \cite{GT}, and its interpretation in terms of the 
AdS/CFT-correspondence \cite{Gubser,GKR,HHR1}. 
 
In the holographic formalism of the AdS/CFT-correspondence, $\phi(x)$  
roughly turns out to be the value of $\Phi(x,y)$ at the boundary,  
$\phi(x)=\Phi(x,y_{\rm bound})$\footnote{Here we disregard the  
subtleties of this identification associated with the asymptotic  
properties of the AdS-spacetime boundary and the conformal structure  
of bulk and brane operators, see \cite{BalGiLa,BalKraLa}.}. In the  
braneworld Randall-Sundrum model, where 
the boundary is associated with the brane $\Sigma$ located at $y_{\rm brane}$, 
this identification has led to the understanding that the recently observed 
localization of the graviton zero mode \cite{RSloc} and the recovery of the 
four-dimensional Einstein gravity on the brane \cite{RSloc,GT,ChGR} can be 
interpreted in terms of the AdS/CFT-correspondence \cite{Gubser,GKR,HHR1}. 
This conclusion was reached in the language of the effective action of the 
brane field --- the four-dimensional metric induced on the brane from the bulk 
geometry. Thus, generically, a natural variable to describe a braneworld 
scenario becomes the value of the field at the brane in question, 
$\phi(x)=\Phi(x,y_{\rm brane})$. Unlike in a Kaluza-Klein reduction, its 
effective action $S_{\rm eff}[\,\phi(x)\,]$ is obtained from the fundamental 
action $S[\,\Phi\,]$ by a less trivial procedure --- by substituting in 
$S[\,\Phi\,]$ a solution of the classical equations of motion for $\Phi(x,y)$ 
in the bulk, $\Phi=\Phi[\,\phi(x)\,]$, parametrized by their boundary values 
on the branes, that is $S_{\rm eff}[\,\phi(x)\,] 
=S\big[\,\Phi[\phi(x)\,]\,\big]$. This construction obviously generalizes to 
the case of several branes $\Sigma_I$ enumerated by the index $I$ and the set 
of brane fields $\phi^I=\Phi(\Sigma_I)$. 
 
Such a definition corresponds to the tree-level approximation for 
the quantum effective action 
    \begin{equation} 
    \exp\Big(iS_{\rm eff}[\,\phi\,]\Big)= 
    \left.\int D\Phi \,\exp\Big(iS[\,\Phi\,]\Big)\, 
    \right|_{\,\Phi(\Sigma)\,=\,\phi}\,\, ,         \label{00.2} 
    \end{equation} 
where the functional integration over the bulk fields runs subject to 
the brane boundary conditions. The scope of this formula is very 
large, because it arises in very different contexts. In 
particular, its Euclidean version $(iS\to -S_{\rm Euclid})$ 
underlies the construction of the no-boundary wavefunction in 
quantum cosmology \cite{HH}. Semiclassically, in the braneworld 
scenario, it represents a Hamilton-Jacobi functional, and its 
evolutionary equations of motion in the ``fifth time'' $y$ can be 
interpreted as renormalization-group equations \cite{Verlinde}. 
It also underlies the effective action formulation of the 
AdS/CFT-correspondence principle between supergravity theory 
on an $AdS_5\times S^5$ background and the superconformal field 
theory on its infinitely remote boundary 
\cite{AdS/CFT,GKR,HHR1,HHR2,SkendSol}. 
 
Here we construct the braneworld effective action of the above 
type for the two-brane Randall-Sundrum model as a functional of 
two induced metrics $g_{\mu\nu}^\pm(x)$ and radion fields 
$\psi^\pm(x)$ on its branes $\Sigma_\pm$. We obtain it 
perturbatively on the background of the Randall-Sundrum solution 
to quadratic order in brane curvatures $R^\pm_{\mu\nu}$ and 
radions. Current interest in this action can be explained by the 
attempts to solve the hierarchy problem and generate 
cosmological scenarios incorporating the dynamics of either 
colliding \cite{DvTyeSh,Ekpyr,brantibr,cycle} or diverging 
\cite{brane} branes. In particular, here we justify the result for 
the braneworld action previously obtained in \cite{brane} by 
a simplified method disregarding curvature perturbations on the 
invisible brane and, thus, confirm the scenario of \cite{brane}  
corresponding to diverging branes. 
 
Another motivation for the two-field braneworld action comes from 
the papers \cite{GarSas} and, especially \cite{GarPujTan}, 
occupying a somewhat intermediate position between the 
Kaluza-Klein setting and the setting of (\ref{00.2}). The authors 
of \cite{GarSas,GarPujTan} use an ansatz in which the two brane 
metrics $g_{\mu\nu}^\pm(x)$ are conformally equivalent, 
$g_{\mu\nu}^+(x)\sim g_{\mu\nu}^-(x)$, and differ only by the 
conformal (warp) factors at the branes. Thus, the braneworld 
action of \cite{GarPujTan} depends on one metric and two radion 
fields associated with the warp factors. This restriction of the 
total configuration space of the model does not leave room for 
important degrees of freedom which are in the focus of this paper. 
 
The organization of the paper is as follows. In Sec.~\ref{struc} 
we focus on the definition of the effective 
action in a braneworld setup and explain the 
method of its calculation. In Sec.~\ref{answer} we 
present the final result for this action in the two-brane 
Randall-Sundrum model, which we advocate here. This action is 
obtained as a quadratic form in Ricci curvatures and radion fields 
on two branes, which is covariant with respect to two independent 
diffeomorphisms associated with these branes. Sec.~\ref{der} is 
devoted to the derivation of this result. The 
derivation begins with the construction of the 
effective equations of motion for the two-brane Randall-Sundrum 
model. Then we recover the action which 
generates these equations by a variational procedure. 
The nonlocal form factors of the braneworld action are explicitly 
constructed in Sec.~\ref{green}. We show that their zeros generate  
the tower of Kaluza-Klein modes and then consider the form factors  
in the few lowest orders of the derivative ($\Box$) expansion. This  
expansion incorporates the massless (graviton) mode of the  
Kaluza-Klein tower, which is responsible for the recovery of  
the four-dimensional Einstein theory on the positive-tension brane.  
In the same section we consider the limit of large interbrane  
distance which turns out to be the high-energy limit on the  
negative-tension brane and discuss the properties of the relevant 
nonlocal operators. In Sec.~\ref{reduced} we build the reduced  
(one-field) effective action corresponding to the on-shell  
reduction in the sector of fields on the negative-tension brane.  
In this way we confirm the result of \cite{brane}, where the  
braneworld action took the 
form of a Brans-Dicke type theory with the radion field 
non-minimally coupled to curvature. In the limit of large brane 
separation we confirm the realization of the 
AdS/CFT-correspondence principle. We also discuss setting Hartle 
boundary conditions at the AdS horizon and the problem of analytic 
continuation to the Euclidean spacetime. The concluding section 
discusses the transition between the local and 
nonlocal phases of the braneworld action. This transition is  
associated with the renormalization type AdS flow from small to  
large interbrane distances, which is likely to be realized  
dynamically due to a repulsive interbrane potential. We briefly  
discuss possible implications of this transition in the diverging  
branes model of \cite{brane}.

\section{Effective action of brane-localized fields and the  
methods of its calculation\label{struc}} 
 
In the definition (\ref{00.2}) the effective action 
by construction depends on the four-dimensional fields 
associated with brane(s). The number of these 
fields equals the number of branes, geometrically each field being 
carried by one of the branes in the system. In the generalized 
Randall-Sundrum setup, the braneworld effective action is 
generated by the path integral of the type (\ref{00.2}), 
     \begin{eqnarray} 
     \left.\int DG \exp\Big(iS[\,G,g,\phi\,]\Big) 
     \right|_{\,\vphantom{I}^4G(\Sigma)\,=\,g}= 
     \exp\Big(iS_{\rm eff}[\,g,\phi\,]\Big),        \label{0.3} 
     \end{eqnarray} 
where the integration over bulk metrics runs subject to fixed 
induced metrics on the branes --- the arguments of $S_{\rm 
eff}[\,g,\phi\,]$. Here $S[\,G,g,\phi\,]$ is the action of the 
five-dimensional gravitational field with the metric 
$G=G_{AB}(x,y)$, $A=(\mu,5),\,\mu=0,1,2,3$, propagating in the 
bulk spacetime ($x^A=(x,y),\,x=x^\mu,\,x^5=y$), and matter fields 
$\phi$ are confined to the branes $\Sigma_I$ --- four-dimensional 
timelike surfaces embedded in the bulk, 
     \begin{eqnarray} 
     &&S[\,G,g,\phi\,]=S_5[\,G\,] 
     +\sum\limits_{I}\int_{\Sigma_I}\! 
     d^4x\,\left(L_{\rm m}(\phi,\partial\phi,g) 
     -g^{1/2}\sigma_I+\frac1{8\pi G_5}[K]\right), \label{0.1}\\ 
     &&S_5[\,G\,]=\frac1{16\pi G_5} 
     \int\limits_{M^5} d^5x\,G^{1/2} 
     \left(\,\vphantom{I}^5\!R(G)-2\Lambda_5\right).  \label{0.2} 
     \end{eqnarray} 
The branes are enumerated by the index $I$ and carry induced metrics 
$g=g_{\mu\nu}(x)$ and matter field Lagrangians 
$L_{\rm m}(\phi,\partial\phi,g)$. The bulk part of the action contains 
the five-dimensional gravitational and cosmological constants, $G_5$ and 
$\Lambda_5$, while the brane parts have four-dimensional cosmological 
constants $\sigma_I$. The bulk cosmological constant $\Lambda_5$ is negative 
and, therefore, is capable of generating the AdS geometry, while 
the brane cosmological constants play the role of brane tensions 
$\sigma_I$ and, depending on the model, can be of either sign. The 
Einstein-Hilbert bulk action (\ref{0.2}) is accompanied by the 
brane `Gibbons-Hawking' terms containing the jump of of the 
extrinsic curvature trace $[K]$ associated with both sides of each brane 
\cite{ReallGH}\footnote{The extrinsic curvature $K_{\mu\nu}$ is 
defined as a projection on the brane of the tensor $\nabla_A n_B$ 
with the outward unit normal $n_B$, i.e. the normal pointing from 
the bulk to the brane. With this definition the normals on the two 
sides of the brane are oppositely oriented and the extrinsic 
curvature jump $[K_{\mu\nu}]$ actually equals the sum of the 
so-defined curvatures on both sides of the brane.}. 
 
In the tree-level approximation the path integral (\ref{0.3}) is 
dominated by the stationary point of the action (\ref{0.1}). Its 
variation is given as a sum of five- and four-dimensional 
integrals, 
    \begin{eqnarray} 
    &&\delta S[\,G,g,\phi\,]=-\frac1{16\pi G_5}\int d^4x\,dy\, 
    G^{1/2}\left(\vphantom{I}^5\!R^{AB} 
     -\frac12\,\vphantom{I}^5\!R\, 
     G^{AB}+\Lambda_5 G^{AB}\right)\delta G_{AB}(x,y)  \nonumber\\ 
     &&\qquad\quad 
     +\sum\limits_{I}\int_{\Sigma_I}\! 
     d^4x\, 
     g^{1/2}\left(-\frac1{16\pi G_5}[K^{\mu\nu} 
     -g^{\mu\nu}K] 
     +\frac12(T^{\mu\nu} 
     -g^{\mu\nu}\sigma)\right)\delta g_{\mu\nu}(x), \label{0.3aa} 
    \end{eqnarray} 
where $\left[K^{\mu\nu}-g^{\mu\nu}K\right]$ denotes the jump of 
the extrinsic curvature terms across the brane, and 
$T^{\mu\nu}(x)$ is the corresponding four-dimensional stress-energy 
tensor of matter fields on the branes, 
    \begin{eqnarray} 
    &&T^{\mu\nu}(x)=\frac2{g^{1/2}} 
    \frac{\delta S_{\rm m}[g,\phi]} 
    {\delta g_{\mu\nu}(x)},         \label{0.4aa}\\ 
    &&S_{\rm m}[\,g,\phi\,]= 
    \sum\limits_{I}\int_{\Sigma_I}\! 
     d^4x\, L_{\rm m}(\phi,\partial\phi,g)        \label{0.4a} 
     \end{eqnarray} 
(we use a collective notation $g$, $\phi$ and $\sigma$ for the induced 
metrics, matter fields and tensions on all branes $\Sigma$). The 
action is stationary when the integrands of both integrals in 
(\ref{0.3aa}) vanish, which gives rise to Einstein equations in 
the bulk, 
    \begin{eqnarray} 
     \frac{\delta S[\,G,g,\phi\,]} 
     {\delta G_{AB}(x,y)}\equiv 
     -\frac1{16\pi G_5} G^{1/2} 
     \left(\vphantom{I}^5R^{AB} 
     -\frac12G^{AB}\,\vphantom{I}^5R 
     +\Lambda_5 G^{AB}\right)=0,   \label{0.6a} 
     \end{eqnarray} 
which are subject to (generalized) Neumann type boundary conditions --- the 
well-known Israel junction conditions --- 
     \begin{eqnarray} 
     \frac{\delta S[\,G,g,\phi\,]} 
     {\delta g_{\mu\nu}(x)} 
     \equiv 
     -\frac1{16\pi G_5}g^{1/2}\left[K^{\mu\nu} 
     -g^{\mu\nu}K\right] 
     +\frac12g^{1/2}(T^{\mu\nu} 
     -g^{\mu\nu}\sigma)=0,              \label{0.7a} 
     \end{eqnarray} 
or to Dirichlet type boundary conditions corresponding to fixed 
(induced) metrics on the branes, with $\delta g_{\mu\nu}=0$ in the 
variation (\ref{0.3aa}), 
     \begin{eqnarray} 
     \vphantom{I}^4G_{\mu\nu}\Big|_\Sigma 
     =g_{\mu\nu}(x)\ .                           \label{0.7} 
     \end{eqnarray} 
 
The solution of the latter, Dirichlet, problem is obviously a 
functional of brane metrics, $G_{AB}=G_{AB}[\,g_{\mu\nu}(x)\,]$, 
and it enters the tree-level approximation for the path integral 
(\ref{0.3}). $S_{\rm eff}[\,g,\phi\,]$ in this approximation 
reduces to the original action (\ref{0.1})--(\ref{0.2}) calculated 
on this solution $G_{AB}[g_{\mu\nu}(x)]$, 
    $S_{\rm eff}[\,g,\phi\,]= 
    S[\,G[\,g\,],g,\phi\,]+O(\hbar)$.  
With this definition, the matter part of effective action 
coincides with the original action Eq.~(\ref{0.4a}) 
     \begin{eqnarray} 
     S_{\rm eff}[\,g,\phi\,]= 
     S_4[\,g\,]+S_{\rm m}[\,g,\phi\,],            \label{0.4} 
     \end{eqnarray} 
while all non-trivial dependence on $g$ arising from the functional 
integration is contained in $S_4[\,g\,]$. 
 
The Dirichlet problem (\ref{0.6a}), (\ref{0.7}) can be regarded as 
an intermediate stage in solving the problem (\ref{0.6a}), 
(\ref{0.7a}). Indeed, given the action (\ref{0.4}) as a result of 
solving the Dirichlet problem (\ref{0.6a}), (\ref{0.7}), one can further 
apply the variational procedure, now with respect to the induced 
metric $g_{\mu\nu}$, to get the effective equations 
     \begin{eqnarray} 
     \frac{\delta S_{\rm eff}[\,g,\phi\,]}{\delta g_{\mu\nu}(x)}= 
     \frac{\delta S_4[\,g\,]}{\delta g_{\mu\nu}(x)} 
     +\frac12 g^{1/2} T^{\mu\nu}(x)=0,              \label{0.8} 
     \end{eqnarray} 
which are equivalent to the Israel junction conditions --- a part 
of the full system of the bulk-brane equations of motion (\ref{0.6a}), 
(\ref{0.7a}) (cf.\ the appendix of \cite{gen}). The procedure of 
solving this system of equations is split into two stages. First 
we solve it in the bulk subject to {\em Dirichlet} boundary 
conditions on the branes and substitute the result into the bulk 
action to get the {\em off-shell} brane effective action. 
Stationarity of the latter with respect to the four-dimensional 
metric comprises the remaining set of equations to be solved at 
the second stage. 
 
This observation suggests two equivalent methods of recovering the braneworld 
effective action. One method is straightforward --- the direct substitution of 
the solution $G_{AB}=G_{AB}[\,g_{\mu\nu}(x)\,]$ of the Dirichlet problem 
(\ref{0.6a}), (\ref{0.7}) into the five-dimensional action. The other method, 
which we choose to pursue in the following, is less direct, but technically is 
simpler. We will recover the effective action from the effective equations 
(\ref{0.8}).  Their left-hand side considered as a variational derivative with 
respect to the brane metric(s) $g_{\mu\nu}$ can be functionally integrated to 
yield $S_{\rm eff}[\,g,\phi]$.  Therefore we will, first, obtain these 
effective equations by solving the bulk part of the equations of motion and by 
explicitly rewriting the Israel matching conditions in terms of $g_{\mu\nu}$. 
The crucial point in the further functional integration of the latter is the 
recovery of a correct integrating factor.  This will be based on the simple 
observation that the stress tensor of matter fields always enters the 
variational derivative of the effective action with the algebraic coefficient 
$(1/2)g^{1/2}$ of Eq.~(\ref{0.8}). In a subsequent publication \cite{duality}  
it will be shown that this derivation of $S_{\rm eff}[\,g,\phi]$ is  
equivalent to 
the first method based on the solution of the Dirichlet boundary value 
problem, (\ref{0.6a}), (\ref{0.7}).

\section{Two-brane Randall-Sundrum model: the final answer 
for the two-field braneworld action\label{answer}} 
 
The action of 
the two-brane Randall-Sundrum model \cite{RS} is given by 
Eq.~(\ref{0.1}) in which the index $I=\pm$ enumerates two branes 
with tensions $\sigma_\pm$. The fifth dimension has the topology 
of a circle labelled by the coordinate $y$, $-d<y\leq d$, with an 
orbifold $Z_2$-identification of points $y$ and $-y$. The branes 
are located at antipodal fixed points of the orbifold, 
$y=y_\pm,\,y_+=0,\,|y_-|=d$. When they are empty, $L_{\rm 
m}(\phi,\partial\phi,g_{\mu\nu})=0$, and their tensions are 
opposite in sign and fine-tuned to the values of $\Lambda_5$ and 
$G_5$, 
     \begin{eqnarray} 
     &&\Lambda_5=-\frac6{l^2},\,\, 
     \sigma_+=-\sigma_-=\frac3{4\pi G_5l},   \label{1.1} 
     \end{eqnarray} 
this model admits a solution with an AdS metric in the 
bulk ($l$ is its curvature radius), 
     \begin{eqnarray} 
     ds^2=dy^2+e^{-2|y|/l}\eta_{\mu\nu}dx^\mu dx^\nu,  \label{1.2} 
     \end{eqnarray} 
$0=y_+\leq|y|\leq y_-=d$, and with a flat induced metric 
$\eta_{\mu\nu}$ on both branes \cite{RS}. The metric on the 
negative tension brane is rescaled by the 
warp factor $\exp(-2d/l)$ providing a possible 
solution for the hierarchy problem \cite{RS}. With the fine tuning 
(\ref{1.1}) this solution exists for arbitrary brane separation 
$d$ --- two flat branes stay in equilibrium. Their flatness is the 
result of compensation between the bulk cosmological constant and 
brane tensions. 
 
Now consider the Randall-Sundrum model with small matter sources 
for metric perturbations $h_{AB}(x,y)$ on the background of this 
solution \cite{RSloc,GT,GKR,ChGR}, 
     \begin{eqnarray} 
     ds^2=dy^2+e^{-2|y|/l}\eta_{\mu\nu}dx^\mu dx^\nu 
     +h_{AB}(x,y)\,dx^Adx^B,                               \label{1.3} 
     \end{eqnarray} 
such that this five-dimensional metric {\em induces} on the branes two 
four-dimensional metrics of the form 
    \begin{eqnarray} 
    g^\pm_{\mu\nu}(x)= 
    a^2_\pm\,\eta_{\mu\nu}+h^\pm_{\mu\nu}(x).  \label{1.4} 
    \end{eqnarray} 
Here the scale factors $a_\pm=a(y_\pm)$ can be expressed in terms of the 
interbrane distance 
    \begin{eqnarray} 
    a_+=1,\,\,\,a_-=e^{-2d/l}\equiv a\ ,       \label{1.5} 
    \end{eqnarray} 
and $h^\pm_{\mu\nu}(x)$ are the perturbations by which the brane 
metrics $g^\pm_{\mu\nu}(x)$ differ from the (conformally) flat 
metrics of the Randall-Sundrum solution (\ref{1.2})\footnote{It is 
needless to emphasize that $h^\pm_{\mu\nu}(x)\neq 
h_{\mu\nu}(x,y_\pm)$ because the induced brane metrics are 
non-trivially related to (\ref{1.3}) via brane embedding 
functions.}. 
 
The main result of this paper is the braneworld effective action 
(\ref{0.4}) calculated for the boundary conditions (\ref{0.7}) of 
this perturbed form (\ref{1.4}). We calculate it in the 
approximation quadratic in perturbations, so that it represents 
the quadratic form in terms of the two-dimensional columns of 
fields $h^\pm_{\mu\nu}(x)$, 
  \begin{equation} 
  {\bf h}_{\mu\nu}=\left[\begin{array}{c} 
  \,h_{\mu\nu}^+(x)\, \\ \, h_{\mu\nu}^-(x)\, 
  \end{array}\right].                        \label{1.6} 
  \end{equation} 
It should be emphasized here that, in contrast to \cite{GarPujTan} 
and other papers on two-brane scenarios, the metric perturbations 
$h_{\mu\nu}^+(x)$ and $h_{\mu\nu}^-(x)$ are independent, which 
makes the configuration space of the theory much richer and 
results in additional degrees of freedom responsible for 
interbrane interaction. 
 
On the other hand, the braneworld effective action is invariant under the 
four-dimensional diffeomorphisms acting on the branes. In the linearized 
approximation they reduce to the transformations of metric perturbations, 
   \begin{equation} 
   h^\pm_{\mu\nu}\rightarrow h^\pm_{\mu\nu} 
   +f^\pm_{\mu\,,\,\nu}+f^\pm_{\nu\,,\,\mu}    \label{1.7} 
   \end{equation} 
with two {\em independent} local vector field parameters 
$f_\mu^\pm=f_\mu^\pm(x)$. Therefore, rather than in terms of 
metric perturbations themselves, the action in question is 
expressible in terms of the tensor invariants of these 
transformations --- linearized Ricci tensors of 
$h_{\mu\nu}=h^\pm_{\mu\nu}(x)$, 
  \begin{eqnarray} 
  R_{\mu\nu}=\frac12\left(-\Box h_{\mu\nu} 
  +h^\lambda_{\nu,\lambda\mu} 
  +h^\lambda_{\mu,\lambda\nu}-h_{,\mu\nu}\right),  \label{1.8} 
  \end{eqnarray} 
on {\em flat} four-dimensional backgrounds of both branes\footnote{Strictly 
speaking, $R^-_{\mu\nu}$ is the linearized Ricci tensor of the artificial 
metric $\eta_{\mu\nu}+h^-_{\mu\nu}$. It differs from the linearized Ricci 
curvature of the second brane, $R_{\mu\nu}(a^2\eta+h^-)=R_{\mu\nu}^-/a^2$, 
by a factor of $a^2$.}. Commas denote partial derivatives, raising and 
lowering of braneworld indices here and everywhere throughout the paper 
is performed with the aid of the flat four-dimensional metric $\eta_{\mu\nu}$, 
   $h^\lambda_\nu\equiv 
   \eta^{\lambda\sigma}h_{\sigma\nu}$, 
   $h\equiv\eta^{\mu\nu}h_{\mu\nu}$, 
   $R=\eta^{\mu\nu}R_{\mu\nu}$, 
and $\Box$ denotes the flat spacetime d'Alembertian 
    \begin{equation} 
   \Box=\eta^{\mu\nu}\partial_\mu\partial_\nu.     \label{1.14} 
   \end{equation} 
 
Finally, we have to describe the variables which determine the 
embedding of branes into the bulk. Due to metric perturbations the 
branes no longer stay at fixed values of the fifth coordinate. Up 
to four-dimensional diffeomorphisms (\ref{1.7}), their embedding 
variables consist of two four-dimensional scalar fields --- the 
radions $\psi^\pm(x)$ --- and, according to the mechanism 
discussed in \cite{brane}, the braneworld action can depend on 
these scalars. Their geometrically invariant meaning is revealed 
in a special coordinate system where the bulk metric perturbations 
$h_{AB}(x,y)$ of Eq.~(\ref{1.3}) satisfy the so called 
Randall-Sundrum gauge conditions, $h_{A5}=0$, 
${h_{\mu\nu}}^{,\,\nu}=h_\mu^\mu=0$. In this coordinate system the 
brane embeddings are defined by the equations 
   \begin{eqnarray} 
   \Sigma_\pm:\,\,\, 
   y=y_\pm+\frac l{a^2_\pm}\psi^\pm(x),\,\,\, 
   y_+=0,\,y_-=d.                          \label{1.11} 
   \end{eqnarray} 
In the approximation linear in perturbation fields and vector 
gauge parameters (in this approximation all these quantities are 
of the same order of magnitude, 
$h^\pm_{\mu\nu}(x)\sim\psi^\pm(x)\sim f^\pm_\mu(x)$), these radion 
fields are invariant under the action of diffeomorphisms 
(\ref{1.7}). 
 
The answer for the braneworld effective action, which we advocate here, 
and which will be derived in the following two sections, 
is given in terms of the invariant fields of the above type, 
$(R^\pm_{\mu\nu}(x),\psi^\pm(x))$, by the following spacetime 
integral of a $2\times2$ quadratic form, 
   \begin{eqnarray} 
   &&S_4\,[\,g^\pm_{\mu\nu},\psi^\pm] 
   =\frac1{16\pi G_4}\int d^4x \left[\,{\bf R}^T_{\mu\nu}\, 
   \frac{2{\bf F}(\Box)}{l^2\Box^2}\,{\bf R}^{\mu\nu} 
   +\frac16\,{\bf R}^{T}\, 
   \frac{{\bf K}(\Box)\!-\!6{\bf F}(\Box)}{l^2\Box^2} 
   \,{\bf R}\right.\nonumber\\ 
   &&\qquad\qquad\qquad\qquad\qquad\qquad\qquad\qquad\qquad 
   \left.-3\Big(\Box{\bf\Psi}\!+\!\frac16{\bf R}\Big)^{\!T}\, 
   \frac{{\bf K}(\Box)}{l^2\Box^2}\, 
   \Big(\Box{\bf\Psi}\!+\!\frac16{\bf R}\Big)\,\right].      \label{1.12} 
   \end{eqnarray} 
Here $G_4$ is an effective four-dimensional gravitational coupling 
constant, 
    \begin{equation} 
    G_4=\frac{G_5}l,                     \label{1.12a} 
    \end{equation} 
$({\bf R}^{\mu\nu},{\bf\Psi})$ and $({\bf 
R}^T_{\mu\nu},{\bf\Psi}^T)$ are the two-dimensional columns 
  \begin{eqnarray} 
   {\bf R}_{\mu\nu} = 
   \left[\begin{array}{c} 
      \,R_{\mu\nu}^+(x)\, \\ \,R_{\mu\nu}^-(x)\, 
  \end{array}\right],\,\,\,\,\, 
  {\bf\Psi}=\left[\begin{array}{c} 
      \,\psi^+(x)\, \\ \, \psi^-(x)\, 
  \end{array}\right]                       \label{1.13a} 
   \end{eqnarray} 
and rows 
   \begin{eqnarray} 
   {\bf R}^T_{\mu\nu}= 
   \Big[\,R^+_{\mu\nu}(x)\,\,\, 
   R^-_{\mu\nu}(x)\,\Big],\,\,\,\,\, 
   {\bf\Psi}^T=\Big[\,\psi^+(x) 
   \,\,\,\psi^-(x)\,\Big],                   \label{1.13} 
  \end{eqnarray} 
of two sets of curvature perturbations and radion fields, 
associated with two branes ($T$ denotes the matrix and vector 
transposition). The indices here are raised as above by the flat 
spacetime metric $\eta_{\mu\nu}$ and ${\bf 
R}\equiv\eta^{\mu\nu}{\bf R}_{\mu\nu}$. The kernels of the quadratic 
forms in (\ref{1.12}) are nonlocal operators --- non-polynomial 
functions of the flat-space d'Alembertian (\ref{1.14}). They are 
both built in terms of the fundamental $2\times2$ matrix-valued 
operator 
  \begin{equation} 
  {\bf F}(\Box)= 
  \left[\begin{array}{cc} 
  \,F_{++}(\Box)\, &\,\,F_{+-}(\Box)\,\\ 
  \,F_{-+}(\Box)\, &\,\,F_{--}(\Box)\, 
  \end{array}\right]                      \label{1.15} 
    \end{equation} 
and powers of the D'Alembertian\footnote{Nonlocalities 
require the prescription of boundary conditions 
which depend on the type of physical problem one is solving. 
We will assume 
that they are specified by a particular type of analytic continuation 
from the Euclidean spacetime in which these boundary conditions are 
trivial --- Dirichlet boundary conditions at the Euclidean braneworld 
infinity, $|x|\to\infty$. This continuation will be discussed below in 
Sec.~\ref{hartle}.}.  In particular, the operator ${\bf K}(\Box)$ reads 
  \begin{equation} 
  {\bf K}(\Box)=2\,{\bf F}(\Box) 
  +\left[\begin{array}{cc} 
      \,-1\,\,&\,0  \\ \,\,0\,\,&\,1/a^2 
  \end{array}\right]\,l^2\Box.               \label{1.16} 
  \end{equation} 
The fundamental operator ${\bf F}(\Box)$ is the inverse of the 
operator-valued matrix ${\bf G}(\Box)$, 
    \begin{equation} 
    {\bf F}(\Box){\bf G}(\Box)={\bf I},        \label{2.27} 
    \end{equation} 
which is determined by the Green's function of the following 
five-dimensional differential operator with Neumann boundary 
conditions, 
    \begin{eqnarray} 
    &&\left(\,\frac{d^2}{d y^2} 
    -\frac{4}{l^2} + 
    \frac{\Box}{a^2 (y)}\,\right) 
    \,G(x,y\,|\,x',y') 
    =\delta^{(4)}(x,x')\,\delta(y-y'),            \label{2.16a}\\ 
    &&\Big(\,\frac d{dy} 
    + \frac{2}{l}\,\Big)\, 
    G(x,y\,|\,x',y')\,\Big|_{\,y=y_\pm}=0.         \label{2.16} 
    \end{eqnarray} 
The kernel of this Green's function rewritten in $x$-space as the 
operator function of $\Box$ parametrically depends on $y$ and $y'$, 
    \begin{equation} 
    G(x,y\,|\,x',y')= 
    l\,G(y,y'|\,\Box)\,\delta(x,x').     \label{2.18} 
    \end{equation} 
Then, the dimensionless elements of the matrix ${\bf 
G}(\Box)=G_{IJ}(\Box)$, $I,J=\pm$, in (\ref{2.27}) are 
    \begin{equation} 
    G_{IJ}(\Box)=G(y_I,y_J|\,\Box).     \label{2.18a} 
    \end{equation} 
 
In the next two sections we derive the braneworld 
effective action (\ref{1.12}) and then 
consider its interpretation and applications in 
braneworld physics. In particular, the concrete form 
of the fundamental operator ${\bf F}(\Box)$ will be given 
in Sec.~\ref{green}--\ref{reduced} where various energy 
limits of this nonlocal form factor will be considered 
in much detail\footnote{Eq.~(\ref{1.12}) gives the action in the 
approximation quadratic in curvature perturbations and radion 
fields $({\bf R}_{\mu\nu},{\bf\Psi})$ with all higher 
order terms discarded. In this approximation it 
is sufficient to keep the nonlocal form factors of the quadratic 
form in (\ref{1.12}) as flat-space ones and build the 
d'Alembertian in cartesian coordinates in terms of partial 
derivatives. In particular, in these coordinates there is no need 
to write the covariant density weights of $g^{1/2}(x)$ 
which become non-trivial only in higher orders of curvature 
expansion. In an arbitrary curvilinear coordinate system the 
expression (\ref{1.12}) should be appropriately covariantized  
by the technique of the covariant perturbation  
theory of \cite{CPT,CPTIII}.}. 
 scite
\section{Derivation of the braneworld effective action 
\label{der}} 
 
\subsection{The effective equations of motion\label{eom}} 
 
The effective equations, as variations of the 
action with respect to brane metrics, see (\ref{0.8}), are 
equivalent to Israel junction conditions with the five-dimensional 
metric coefficients $G_{AB}(x,y)$ (entering extrinsic curvatures 
of the branes) expressed in terms of $g_{\mu\nu}^\pm(x)$ and 
$T^{\mu\nu\,\pm}(x)$. So to disentangle effective four-dimensional 
equations we have to solve the bulk equations of motion for 
$G_{AB}(x,y)$ in terms of $g_{\mu\nu}^\pm(x)$ and 
$T^{\mu\nu\,\pm}(x)$ and substitute the result in the junction 
conditions. Implicitly this has been done at the exact level in 
\cite{ShiMaSa} where all unknown terms were isolated in form of 
the five-dimensional Weyl tensor and its derivatives. For the 
linearized theory this can be done explicitly along the lines of 
\cite{GT}. 
 
We start by linearizing the five-dimensional Einstein equations 
(\ref{0.6a}) in terms of metric perturbations $h_{AB}(x,y)$ on the 
background of the Randall-Sundrum solution (\ref{1.3}). 
One can always go to the coordinate system with 
    $h_{55}= h_{\mu 5} = 0$, 
    $h^\nu_{\mu\,,\,\nu} =0$, 
    $h^{\mu}{}_{\mu} = 0$ 
\cite{ChGR,GKR}\footnote{Usually this is called 
the Randall-Sundrum gauge. However, this is a combination of the 
gauge-fixing procedure and the use of the non-dynamical (constraint) 
part of linearized Einstein equations in the bulk.}, 
in which the linearized equations in the bulk simplify to a 
single equation for transverse-traceless $h_{\mu\nu}$, 
    \begin{eqnarray} 
    \left(\frac{d^2}{d y^2} 
    -\frac{4}{l^2} + \frac{\Box}{a^2 (y)}\right) 
    \,h_{\mu \nu}(x,y) = 0,     \label{2.11} 
    \end{eqnarray} 
 
The Randall-Sundrum coordinate system is, however, not 
Gaussian normal relative to $\Sigma_\pm$ --- both branes in these 
coordinates are {\em not} located at constant values of the fifth 
coordinate $y_\pm$. Therefore, let us introduce two coordinate systems 
which are Gaussian normal relative to their respective branes and 
mark them by $(\pm)$ \cite{ChGR}. Metric perturbations in these 
coordinate systems, denoted correspondingly by 
$h_{\mu\nu}^{(\pm)}(x,y)$, also have only $\mu\nu$-components, 
$\Sigma_+$ is located at constant value $y_+$ in the 
$(+)$-coordinates, while $\Sigma_-$ has in these coordinates a 
non-trivial embedding. Vice versa, in $(-)$-coordinates $\Sigma_-$ 
is located at fixed value $y_-$, while $\Sigma_+$ is embedded 
non-trivially. Metric perturbations $h_{\mu\nu}^{(\pm)}(x,y)$ in 
$(\pm)$-coordinates are related to those of the Randall-Sundrum 
coordinate system, $h_{\mu\nu}(x,y)$, by remnant coordinate 
transformations --- the transformations that preserve the 
conditions $h_{5A}(x,y)=0$. Every such transformation is 
parametrized by one four-dimensional scalar field $\xi(x)$ and one 
four-dimensional vector field $\xi_\mu(x)$ and has a fixed 
dependence on the fifth coordinate \cite{GT}. These 
transformations to two Gaussian normal coordinate systems read 
    \begin{equation} 
    h^{(\pm)}_{\mu \nu}(x,y) 
    = h_{\mu \nu}(x,y)+ l \xi^\pm_{,\,\mu \nu}(x) 
    +\frac{2}{l} \eta_{\mu \nu} a^2(y)\,\xi^\pm(x) 
    +a^2(y)\,\xi^\pm_{(\mu,\nu)}(x)                \label{2.13} 
    \end{equation} 
and, thus, give rise to two radion fields $\xi^\pm(x)$ and two 
four-dimensional vector fields $\xi^\pm_\mu(x)$. 
 
In the $(\pm)$ Gaussian normal coordinates, five-dimensional 
metrics $g^{(\pm)}_{\mu\nu}(x,y)$ are given by Eq.~(\ref{1.3}) with 
perturbations $h^{(\pm)}_{\mu\nu}(x,y)$, 
    \begin{eqnarray} 
     ds^2=dy^2+e^{-2|y|/l}\eta_{\mu\nu}dx^\mu dx^\nu 
     +h^{(\pm)}_{\mu\nu}(x,y)\,dx^\mu dx^\nu.          \label{1.3a} 
     \end{eqnarray} 
The extrinsic curvatures of $\Sigma_\pm$ in these coordinates simplify 
to 
    $K_{\mu\nu}^\pm=\mp{1\over2} 
    dg^{(\pm)}_{\mu\nu}(x,y)/dy\,|_{\,y=y_\pm}$, 
where the sign factor originates from different orientation of 
outward normals to the corresponding branes. Therefore, the linearized 
boundary conditions (\ref{0.7a}) take the form 
    \begin{equation} 
    \Big(\,\frac d{dy} + \frac{2}{l}\,\Big) 
    h_{\mu \nu}^{(\pm)}(x,y)\Big|_{\,y=y_\pm} 
    =\mp 8\pi G_5 
    \Big(T_{\mu\nu}-\frac13\eta_{\mu \nu}T\Big)^\pm, \label{2.14} 
    \end{equation} 
where\footnote{In the definition of $T_{\mu\nu}$ we have to 
deviate from the general rule of lowering the indices with the 
flat metric $\eta_{\mu\nu}$ because in the exact theory, the covariant 
stress tensor is related to the contravariant one via the full 
metric $g_{\mu\nu}^\pm=a^2_\pm\eta_{\mu\nu}+O(h_{\mu\nu})$.} 
    \begin{eqnarray} 
    T^\pm_{\mu\nu}\equiv 
    a_\pm^4\eta_{\mu\alpha}\eta_{\nu\beta} 
    T^{\alpha\beta}_\pm= 
    2\eta_{\mu\alpha}\eta_{\nu\beta} 
    \frac{\delta 
    S_{\rm m}}{\delta h^\pm_{\alpha\beta}},\,\,\,\, 
    T^\pm\equiv\eta^{\mu\nu}T^\pm_{\mu\nu}. \label{2.14a} 
    \end{eqnarray} 
Here we took into account that in view of the orbifold 
$Z_2$-symmetry the curvature jump is 
$[K_{\mu\nu}]^\pm=2K_{\mu\nu}^\pm$. Using (\ref{2.13}) in 
(\ref{2.14}) and taking into account that $(d/dy + 2/l) a^2(y)=0$ 
we finally find 
    \begin{eqnarray} 
    \Big(\,\frac d{dy}+\frac{2}{l}\,\Big) 
    h_{\mu \nu}(x,y)\Big|_{\,y=y_\pm} = 
    \mp 8\pi G_5 \Big( T_{\mu \nu} - 
    \frac13 \eta_{\mu \nu} 
    T \Big)^\pm-2 \xi^\pm_{,\,\mu \nu}.               \label{2.15} 
    \end{eqnarray} 
These are the linear boundary conditions on perturbations $h_{\mu 
\nu}(x,y)$ in the Randall-Sundrum gauge\footnote{Strictly 
speaking, in the Randall-Sundrum coordinate system the branes are 
shifted from the constant values $y_\pm$. However, this displacement 
is of first order of magnitude in $\xi^\pm(x)\sim 
h^\pm_{\mu\nu}(x)$ and, therefore, it contributes to (\ref{2.15}) 
only in higher orders of perturbation theory.}. 
 
By taking the trace of this equation one finds on account of 
tracelessness of $h_{\mu\nu}$ the dynamical four-dimensional 
equation for radion fields \cite{GT}, 
    \begin{eqnarray} 
   \Box \xi^\pm(x) =\pm\frac{8\pi G_5}6 T^\pm(x).  \label{2.15a} 
    \end{eqnarray} 
 
Now the system of equations and the boundary conditions becomes 
complete. 
Eq.~(\ref{2.15a}) determines the radions, while the boundary value 
problem (\ref{2.11}) and (\ref{2.15}) determines the perturbations 
in the bulk. The solution to this problem can be given in terms of 
the Green's function of the equation (\ref{2.11}) with homogeneous 
Neumann boundary conditions. This Green's function 
satisfies (\ref{2.16a})--(\ref{2.16}). The desired solution reads 
    \begin{eqnarray} 
    &&h_{\mu\nu}(x,y)=l\,G(y,y_+|\,\Box)\, 
    w^+_{\mu\nu}(x)- 
    l\,G(y,y_-|\,\Box)\,w^-_{\mu\nu}(x),    \label{2.17}\\ 
    &&w^\pm_{\mu\nu}= 
    \mp 8\pi G_5 \left( T_{\mu \nu} - 
    \frac13 \eta_{\mu \nu} 
    T \right)^\pm-2\xi^\pm,_{\mu \nu},  \nonumber 
    \end{eqnarray} 
where for brevity we have denoted the right-hand sides of the 
(inhomogeneous) boundary conditions (\ref{2.15}) by 
$w^\pm_{\mu\nu}$ and used the shorthand notation for the kernel of 
the five-dimensional Green's function (\ref{2.18}), which allows 
one to omit the $x$-integration signs. 
 
We will be interested in the effective dynamics of the 
observable fields only --- the induced metric perturbations on the 
branes $h_{\mu\nu}^\pm(x)$. In the notation of Sec.~\ref{answer} 
they form the column (\ref{1.6}), and the Green's function of 
Eq.~(\ref{2.17}) can be regarded as the $2\times2$-matrix 
(\ref{2.18a}), acting in the space of such columns. With these 
notations the observable metric perturbations in the 
Randall-Sundrum gauge read as the following (nonlocal) linear 
combination of stress tensors and radion fields, 
    \begin{equation} 
    \left[\begin{array}{c} 
    \,h_{\mu\nu}(x,y_+)\, \\ \, h_{\mu\nu}(x,y_-)\, 
    \end{array}\right]= 
    - 8 \pi G_5 l\,{\bf G}(\Box) 
    \left[\begin{array}{c} 
    \,T^+_{\mu \nu} - \frac{1}{3} \eta_{\mu \nu} T^+\, \\ 
      \,T^-_{\mu \nu} - \frac{1}{3} \eta_{\mu \nu} T^-\, 
     \end{array}\right] 
   - 2l\,{\bf G}(\Box) \left[\begin{array}{c} 
       \,\xi^+,_{\mu \nu}\, \\ 
                 \, - \xi^-,_{\mu \nu} 
      \end{array}\right].                    \label{2.21} 
    \end{equation} 
These perturbations do not, however, coincide with those of the 
{\em induced} metrics on branes, $h_{\mu\nu}^\pm$. The latter 
coincide with the metric coefficients in two respective Gaussian 
normal coordinate systems, $h_{\mu\nu}^\pm(x)\equiv 
h^{(\pm)}_{\mu\nu}(x,y_\pm)$. Therefore, $h_{\mu\nu}^\pm(x)$ is 
related to $h_{\mu\nu}(x,y_\pm)$ by Eq.~(\ref{2.13}) and read in 
column notations of Eq.(\ref{1.6}) 
    \begin{eqnarray} 
    &&{\bf h}_{\mu\nu}(x)= 
  - 8 \pi G_5 l\,{\bf G}(\Box)\, 
      \Big(\,{\bf T}_{\mu \nu} 
      - \frac{1}{3} \eta_{\mu \nu} {\bf T}\,\Big) 
   +\frac2l\eta_{\mu\nu}\left[\begin{array}{c} 
      \,\xi^+\, \\ \, a^2\xi^-\, \end{array}\right]\nonumber\\ 
    &&\qquad\qquad\qquad\qquad\qquad\quad 
    +l\left(\left[\begin{array}{cr} 
      \,1\,&\,0\, \\ \,0\,&\,-1\, \end{array}\right] 
    - 2{\bf G}(\Box)\right) \left[\begin{array}{c} 
       \,\xi^+_{,\mu \nu}\,\\ 
       \,- \xi^-_{,\mu \nu}\end{array}\right] 
   +\left[\begin{array}{c} 
      \,\xi^+_{(\mu,\nu)}\, 
      \\ \, a^2\xi^-_{(\mu,\nu)}\, \end{array}\right], \label{2.22} 
    \end{eqnarray} 
where we took into account that $a_+=1$ and $a_-\equiv a$, 
Eq.~(\ref{1.5}). 
 
 From the four-dimensional viewpoint, the last two terms in this 
expression represent two diffeomorphisms on the branes that can 
be gauged away, so that the induced metric perturbations (we denote 
them in the new gauge by $H^\pm_{\mu\nu}$) finally read 
    \begin{eqnarray} 
    {\bf H}_{\mu\nu}= 
    - 8 \pi G_5 l\,{\bf G}(\Box) 
    \Big(\,{\bf T}_{\mu \nu} 
      - \frac{1}{3} \eta_{\mu \nu} {\bf T}\,\Big) 
    +\frac2l\eta_{\mu\nu}\left[\begin{array}{c} 
      \,\xi^+\, \\ \, a^2\xi^-\, \end{array}\right].  \label{2.23} 
    \end{eqnarray} 
The gauge conditions for $H_{\mu\nu}^\pm(x)$ are unknown and 
differ from those of $h_{\mu\nu}^\pm(x)$ (also unknown ones), but 
they will be determined later when covariantizing the effective 
equations of motion in terms of four-dimensional curvatures. 
 
In Eq.~(\ref{2.23}) we have a typical five-dimensional combination of 
the stress tensor and its trace, 
$T_{\mu\nu}-{1\over3}\eta_{\mu\nu}T$. This can be rewritten as the 
four-dimensional combination $T_{\mu\nu}-{1\over2}\eta_{\mu\nu}T$ 
plus the contribution of the trace ${1\over6}\eta_{\mu\nu}T$ which 
can be expressed in terms of $\xi$ according to (\ref{2.15a}). 
Then Eq.~(\ref{2.23}) takes the form 
    \begin{eqnarray} 
    {\bf H}_{\mu\nu}= 
    - 8 \pi G_5 l\,{\bf G}(\Box) 
    \Big(\,{\bf T}_{\mu \nu} 
      - \frac{1}{2} \eta_{\mu \nu} {\bf T}\,\Big) 
    +l\eta_{\mu\nu}\Box{\bf G_\xi} 
    (\Box)\left[\begin{array}{c} 
    \,\xi^+\, \\ \,\xi^-\, \end{array}\right],  \label{2.23a} 
    \end{eqnarray} 
where the new operator ${\bf G_\xi}(\Box)$ reads in terms of the 
brane separation parameter $a$ as 
    \begin{eqnarray} 
    {\bf G_\xi}(\Box) \equiv {\bf G}(\Box) 
    \left[\begin{array}{cc} 
    \,-1\,\,&\,0\,\\ \,0\,&\,1\,\end{array}\right] 
    + \frac{2}{l^2\, 
    \Box}\left[\begin{array}{cc} 
    \,1\,&\,0\,\\ \,0\,&\,a^2\,\end{array}\right].  \label{2.24} 
    \end{eqnarray} 
As we will see below, such a rearrangement is very illuminating 
when recovering the effective four-dimensional Einstein theory in the 
low-energy approximation. 
 
The next step consists in the covariantization of these equations. 
Thus far they are written in terms of brane metric perturbations 
$h_{\mu\nu}^\pm=H^\pm_{\mu\nu}$ in a particular gauge 
corresponding to omission of gauge transformation terms on the 
right-hand side of (\ref{2.22}). It is always useful to have the 
dynamical equations in gauge independent form. This form can be 
easily attained by rewriting them in terms of the curvature. For 
this purpose we first determine explicitly the gauge conditions 
for the perturbations $H^\pm_{\mu\nu}$ and then express these 
perturbations in terms of the linearized Ricci tensors of brane 
metrics. The gauge for $H^\pm_{\mu\nu}$ can be found by going back 
to the original transverse-traceless perturbations 
$h_{\mu\nu}^\pm(x,y_\pm)$. However, it is much easier to recover 
these gauge conditions directly from the equations of motion 
(\ref{2.23}). By applying the first-order differential operator of 
harmonic gauge conditions to $H^\pm_{\mu\nu}$ in (\ref{2.23}) and 
using the conservation law for matter stress tensors we find that 
    \begin{equation} 
    {\bf H}^{\nu}_{\mu\,,\,\nu}- 
      \frac12\,{\bf H}_{,\,\mu} 
      =-l\Box{\bf G_\xi}(\Box) 
    \left[\begin{array}{c} 
      \,\xi^+_{,\,\mu}\,\\ 
      \,\xi^-_{,\,\mu}\,\end{array}\right].  \label{2.24a} 
    \end{equation} 
These equations serve as a set of generalized harmonic gauge 
conditions on brane metrics $H^\pm_{\mu\nu}$ and radions 
$\xi^\pm$. Now, one can use them in Eq.~(\ref{1.8}) for the 
linearized Ricci tensor in order to obtain the equation for 
$H_{\mu\nu}$ that can be solved by iterations in powers of 
$R_{\mu\nu}$ and $\xi$ \cite{CPT}. In the linear approximation 
this solution reads 
    \begin{equation} 
    {\bf H}_{\mu\nu}=-\frac2\Box {\bf R}_{\mu\nu} 
    -2l\,{\bf G_\xi}(\Box) 
    \left[\begin{array}{c} 
    \,\xi^+_{,\,\mu\nu}\,\\ \,\xi^-_{,\,\mu\nu}\, 
    \end{array}\right].                            \label{2.24b} 
    \end{equation} 
 
Using (\ref{2.24b}) in Eq.~(\ref{2.23}), we rewrite the latter in 
terms of the linearized Einstein tensors 
$E_{\mu\nu}^\pm=R_{\mu\nu}^\pm-{1\over2}\eta_{\mu\nu}R^\pm$ of the 
brane metrics, 
    \begin{equation} 
    -\frac2\Box {\bf E}_{\mu\nu} 
    +8\pi G_5\, l\,{\bf G}(\Box) 
    {\bf T}_{\mu \nu} 
   -2l\,(\nabla_\mu\nabla_\nu-\eta_{\mu\nu}\Box)\,{\bf G_\xi}(\Box) 
   \left[\begin{array}{c} 
      \,\xi^+\, \\ \,\xi^-\, 
      \end{array}\right]=0. \label{2.25} 
    \end{equation} 
Now these equations are covariant and equally valid in terms of 
brane metrics $g^\pm_{\mu\nu}=a_\pm^2\eta_{\mu\nu}+h^\pm_{\mu\nu}$ 
with perturbations $h^\pm_{\mu\nu}$ taken in {\em any} gauge, not 
necessarily coinciding with that of $H^\pm_{\mu\nu}$. 
 
As it was discussed earlier, we have to find the action that 
generates these equations by the variational procedure. For this 
purpose we must find the integrating factor --- the overall matrix 
valued coefficient --- with which the left-hand side of 
(\ref{2.25}) enters the variational derivative of the action, see 
(\ref{0.8}). To find it, we note that matter fields living on 
branes are directly coupled only to 
$g^\pm_{\mu\nu}$ via their stress tensors (\ref{0.4aa}) and, 
moreover, the matter action additively enters the full effective 
action (\ref{0.4}). Therefore, in the linear approximation the 
overall coefficients of $T^{\mu\nu}_\pm$ in the variational 
derivatives of the action should be the local numericals 
${1\over2}a_\pm^4$ (remember that in cartesian 
coordinates $g_\pm^{1/2}=a_\pm^4+O(h_{\mu\nu})$). To 
achieve this we raise the indices in (\ref{2.25}) with 
$\eta^{\mu\nu}$ and act upon it by the operator ${\bf 
F}(\Box)/16\pi G_5l$, where ${\bf F}(\Box)$ is the 
inverse of the Green's function ${\bf G}(\Box)$, Eq.~(\ref{2.27}). 
Then, the left hand side of eq. (\ref{2.25}) as a metric 
variational derivative takes the form (${\bf \delta/\delta 
g_{\mu\nu}}$ denotes the column of derivatives with respect to 
$g^\pm_{\mu\nu}$) 
    \begin{eqnarray} 
    &&{\bf \frac\delta{\delta g_{\mu\nu}}} 
    S_{\rm eff}\,[\,g^\pm,\xi^\pm,\phi^\pm\,]= 
    -\frac1{8\pi G_5l} 
    \frac{{\bf F}(\Box)}{\Box}\, 
    {\bf E}^{\mu\nu}+\frac12 
    {\bf g^{1/2}T}^{\mu \nu}           \nonumber\\ 
    &&\qquad\qquad\qquad\qquad\qquad\qquad\quad-\frac1{8\pi G_5}\, 
    (\nabla^\mu\nabla^\nu-\eta^{\mu\nu}\Box)\, 
    {\bf F}(\Box)\,{\bf G_\xi}(\Box) 
    \left[\begin{array}{c} 
      \,\,\xi^+\, \\ \,\,\xi^-\, \end{array}\right].    \label{2.28} 
    \end{eqnarray} 
Here we take into account that in view of the 
definition (\ref{2.14a}) $\eta^{\mu\alpha}\eta^{\nu\beta} 
T_{\alpha\beta}^\pm=a_\pm^4T^{\mu\nu}_\pm=g^{1/2}_\pm T^{\mu\nu}_\pm$. 
 
Our next goal is to rewrite the equations for the radions 
(\ref{2.15a}) also as variational equations for the braneworld 
effective action. Again, matter cannot be directly coupled to 
radions because the matter action 
additively enters the full action. Therefore, the stress tensor 
part of (\ref{2.15a}) should be reexpressed in terms of the 
curvature. By taking the trace of 
(\ref{2.28}) and using (\ref{2.15a}) one can exclude the stress 
tensor traces in terms of $R^\pm$ and 
$\xi^\pm$, so that the dynamical equations for the latter 
reduce, in view of (\ref{2.24}), to 
    \begin{equation} 
    \left[\begin{array}{c} 
      \,R^+\, \\ 
      \, R^-\, 
    \end{array}\right]+\frac6l\left[\begin{array}{cc} 
      \,1\,\,&\,0\, \\ 
      \, 0\,\,&\,a^2 \, 
    \end{array}\right]\Box \left[\begin{array}{c} 
      \,\,\xi^+\, \\ 
      \,\, \xi^-\, 
    \end{array}\right]=0.                           \label{2.29} 
    \end{equation} 
Actually, Eq.~(\ref{2.29}) has a simple physical interpretation. It is the 
linearised equation of motion for worldsheet perturbations of the brane 
coupled to bulk perturbations \cite{worldsheet}. 
In the next section we recover the braneworld effective action 
that generates the full set of equations (\ref{2.28}) and 
(\ref{2.29}).

\subsection{The recovery of the effective action 
\label{action}} 
 From the structure of (\ref{2.28}) it is obvious that the  
graviton-radion part of the full braneworld action  
$ S_4\,[\,g^\pm_{\mu\nu},\xi^\pm]$ is given 
by the sum of the purely gravitational part, the non-minimal 
coupling of the radions to the curvatures and the radion action itself, 
    \begin{equation} 
    S_4\,[\,g^\pm_{\mu\nu},\xi^\pm] 
    =S_{\rm grav}[\,g^\pm_{\mu\nu}]+ 
    S_{\rm n-m}[\,g^\pm_{\mu\nu},\xi^\pm]+ 
    S_{\rm rad}\,[\,g^\pm_{\mu\nu},\xi^\pm]. 
    \end{equation} 
Here $S_{\rm grav}$ and $S_{\rm n-m}$ give as contributions to 
the variational derivative of the action the first and the third 
terms on the right-hand side of (\ref{2.28}), while 
$S_{\rm rad}[\,g^\pm_{\mu\nu},\xi^\pm]$ still has to be 
found by comparing its radion variational derivative with 
(\ref{2.29}). 
 
Due to the linearity of (\ref{2.29}) in the metric perturbations, $S_{\rm 
  grav}+S_{\rm n-m}$ can be obtained from the right-hand side of (\ref{2.28}) 
by contracting it with the row of metric perturbations ${\bf H}^T_{\mu\nu}$ 
and integrating over $x$. Using (\ref{2.24b}), one can convert the result to 
the final form quadratic in Ricci curvatures 
    \begin{eqnarray} 
    &&S_{\rm grav}[\,g^\pm]=\frac1{8\pi G_5 l}\int d^4x\, 
    {\bf R}_{\mu\nu}^T 
    \frac{{\bf F}(\Box)}{\Box^2}\, 
    {\bf E}^{\mu\nu},                      \label{3.3}\\ 
    &&S_{\rm n-m}[\,g^\pm,\xi^\pm]= 
    -\frac1{8\pi G_5}\,\int d^4x\, 
    \Big[\,\xi^+\,\,\,\xi^-\Big]\, 
   {\bf F}(\Box){\bf G_\xi}(\Box)\,{\bf R}.    \label{3.4} 
    \end{eqnarray} 
With the action (\ref{3.4}) for the non-minimal coupling the equations of 
motion for the radions read 
    \begin{equation} 
    \left[\begin{array}{c} 
      \,\delta/\delta\xi^+\, \\ 
      \, \delta/\delta\xi^-\, 
    \end{array}\right]\Big(S_{\rm rad}+S_{\rm n-m}\Big)= 
    \left[\begin{array}{c} 
      \,\delta/\delta\xi^+\, \\ 
      \, \delta/\delta\xi^-\, 
    \end{array}\right]\,S_{\rm rad} 
    -\frac1{8\pi G_5} 
   {\bf G}_\xi^T(\Box){\bf F}^T(\Box)\,{\bf R}=0. 
    \end{equation} 
The comparison with (\ref{2.29}) shows that 
    \begin{eqnarray} 
    &&\left[\begin{array}{c} 
      \,\delta/\delta\xi^+\, \\ 
      \, \delta/\delta\xi^-\, 
    \end{array}\right]\,S_{\rm rad}=-\frac6{8\pi G_5 l^3} 
    \left[\begin{array}{cc} 
     \, \,1\,\,&\,0\, \\ \,\,0\,\,&\,a^2\, 
    \end{array}\right] 
    {\bf K}(\Box) 
    \left[\begin{array}{cc} 
     \, \,1\,\,&\,0\, \\ \,\,0\,\,&\,a^2\, 
    \end{array}\right]\left[\begin{array}{c} 
      \,\,\xi^+\, \\ \,\,\xi^-\, 
    \end{array}\right],                           \label{3.5}\\ 
    &&{\bf K}(\Box)\equiv 
    \left[\begin{array}{cc} 
     \, \,1\,\,&\,0 \\ \,\,0\,\,&\,1/a^2 
    \end{array}\right]{\bf G}_\xi^T(\Box) 
    {\bf F}^T(\Box)\,l^2\Box\ . 
    \end{eqnarray} 
 
As follows from the definition of ${\bf G}_\xi(\Box)$, 
Eq.~(\ref{2.24}), the matrix ${\bf K}(\Box)$ is given by 
Eq.~(\ref{1.16}). Thus, it is symmetric in view of the symmetry of 
${\bf F}(\Box)$. Eq.~(\ref{3.5}) is therefore integrable 
and the radion action acquires the quadratic form with this new 
matrix valued operator ${\bf K}(\Box)$ as a kernel. The equations 
become simpler if we introduce instead of $\xi^\pm$ the 
new {\em dimensionless} radion fields 
    \begin{eqnarray} 
    \psi^+=\frac{\xi^+}l, \,\,\,\, 
    \psi^-=a^2\,\frac{\xi^-}l,                \label{3.7} 
    \end{eqnarray} 
in terms of which the non-minimal coupling and radion actions read 
    \begin{eqnarray} 
    &&S_{\rm n-m}[\,g^\pm,\xi^\pm]= 
    -\frac l{8\pi G_5}\,\int d^4x\,{\bf R}^T\, 
    \frac{{\bf K}(\Box)}{l^2\Box}\, 
    {\bf\Psi},                           \label{3.8}\\ 
    &&S_{\rm rad}[\,g^\pm,\xi^\pm]= 
    -\frac{3l}{8\pi G_5}\,\int d^4x\, 
    {\bf\Psi}^T\, 
    \frac{{\bf K}(\Box)}{l^2}\, 
    {\bf\Psi}\ .                          \label{3.9} 
    \end{eqnarray} 
 
    Thus, the sum of gravitational (\ref{3.3}), non-minimal (\ref{3.8}) and 
    radion (\ref{3.9}) parts form the full braneworld effective action in the 
    approximation quadratic in fields.  Collecting the radion and non-minimal 
    parts one can easily recover the last term of (\ref{1.12}) --- the 
  $2\times2$ quadratic form in $(\Box{\bf\Psi}+{\bf R}/6)$ --- plus an extra 
    term quadratic in the Ricci scalars on the branes ${\bf R}$. 
When combined with 
    (\ref{3.3}), the latter gives rise to the first two terms of (\ref{1.12}). 
    This accomplishes the derivation of the basic result advocated in 
    Sec.~\ref{answer}. The nonlocal kernels in the quadratic forms of 
    (\ref{1.12}) are built in terms of $2\times2$-matrix-valued operators ${\bf 
      F}(\Box)$ and ${\bf K}(\Box)$ which we analyze in the next section. 
 
\section{Nonlocal form factors of the braneworld action\label{green}} 
 
To construct the operators ${\bf F}(\Box)$ and 
${\bf K}(\Box)$ we need the Green's function of the boundary value 
problem (\ref{2.16a})--(\ref{2.16}). This problem simplifies to 
the Bessel's equation in terms of the new variable $z$, 
    \begin{equation} 
    z=l\exp\frac yl,\,\,\,a(y)=\frac lz.    \label{4.1} 
    \end{equation} 
For the dimensionless function $\bar G(z,z'|\,\Box)\equiv 
G(y,y'|\,\Box)$ defined by Eq.~(\ref{2.18}) it reads as 
    \begin{eqnarray} 
    &&\left(\,\frac d{dz}z\frac d{dz} 
    +z\Box-\frac4z\,\right) 
    \bar G(z,z'|\,\Box)=\delta(z-z'),           \label{4.2} \\ 
    &&\frac d{dz}z^2\bar 
    G(z,z'|\,\Box)\,\Big|_{z=z_\pm}=0.        \label{4.3} 
    \end{eqnarray} 
This Green's function can be built in terms of the basis functions 
$u_\pm(z)=u_\pm(z|\,\Box)$ of Eq.~(\ref{4.2}) --- two linearly 
independent solutions of the homogeneous Bessel equation 
satisfying Neumann boundary conditions at $z_+$ and 
$z_-$, respectively, 
    \begin{eqnarray} 
    &&\left(\,\frac d{dz}z\frac d{dz} 
    +z\Box-\frac4z\,\right)\,u_\pm(z)=0,\,\,\, 
    z_+\leq z\leq z_-,                             \label{4.4} \\ 
    &&\frac d{dz}z^2 u_\pm(z)\,\Big|_{z=z_\pm}=0.  \label{4.5} 
    \end{eqnarray} 
They are given by linear combinations of Bessel and Neumann 
functions of the second order, 
$Z_2(z\sqrt\Box)=(J_2(z\sqrt\Box),Y_2(z\sqrt\Box))$, with the 
coefficients easily derivable from the boundary conditions on account 
of the relation $(d/dx)x^2 Z_2(x)=x^2 Z_1(x)$, 
    \begin{eqnarray} 
    &&u_\pm(z)=Y_1^\pm J_2(z\sqrt\Box) 
    -J_1^\pm Y_2(z\sqrt\Box),                        \label{4.6}\\ 
    &&J_1^\pm \equiv J_1(z_\pm\sqrt\Box),\,\,\,\, 
    Y_1^\pm \equiv Y_1(z_\pm\sqrt\Box). 
    \end{eqnarray} 
In what follows we introduce the abbreviation for the Bessel function of any 
order $Z_\nu^\pm\equiv Z_\nu(z_\pm\sqrt\Box)$ to avoid excessive use of their 
different arguments. 
 
In terms of $u_\pm(z)$, the Green's function has a well known representation, 
    \begin{equation} 
     \bar G(z,z'|\,\Box)=\theta(z-z') 
     \frac{u_-(z)u_+(z')}{\Delta}+ 
    \theta(z'-z)\frac{u_+(z)u_-(z')}{\Delta},              \label{4.7} 
    \end{equation} 
where $\theta(z)$ is the step function, $\theta(z)=1,\,z\geq 0$, 
$\theta(z)=0,\,z<0$, and $\Delta$ is the conserved Wronskian inner product 
of basis functions in the space of solutions of Eq.~(\ref{4.4}), 
    \begin{eqnarray} 
    \Delta\equiv z\Big(u_+(z)\frac d{dz}u_-(z) 
    -u_-(z)\frac d{dz}u_+(z)\Big)= 
    \frac2\pi\,\Big(J_1^+\,Y_1^--Y_1^+J_1^-\Big).  \label{4.8} 
    \end{eqnarray} 
 
The calculation of the $2\times2$-matrix Green's function 
(\ref{2.18a}) on the basis of (\ref{4.7}) requires the knowledge 
of $u_+(z_\pm)$ and $u_-(z_\pm)$. Some of these simplify to 
elementary functions, 
    $u_\pm(z_\pm)=Y_1^\pm J_2^\pm 
    -J_1^\pm Y_2^\pm=2/{\pi z_\pm\sqrt\Box}$. 
Using this result together with (\ref{4.8}) in (\ref{4.7}), one 
finds the following exact expression for ${\bf G}(\Box)$, 
    \begin{equation} 
    {\bf G}(\Box) =\frac{1}{\Box z_+ z_-} 
    \frac{1}{J_1^+\,Y_1^-  - J_1^-\,Y_1^+ } 
    \left[\begin{array}{cc} 
    \,\sqrt{\Box} z_-u_-(z_+)&\displaystyle{\frac2\pi} \\ 
    \displaystyle{\frac2\pi}&\sqrt{\Box} z_+u_+(z_-)\, 
    \end{array}\right].                               \label{4.10} 
    \end{equation} 
%New 
    The Green's functions in ${\bf G}(\Box)$ have first been calculated in 
    \cite{Grinstein}. 
%New 
 
Interestingly, its inverse ${\bf F}(\Box)$ can be represented, in essence, 
in a form dual to this expression. Indeed, for the 
inversion of the matrix (\ref{4.10}) we need its determinant which 
can be shown to equal\footnote{To derive this representation for 
the determinant one should explicitly use the bilinear in Bessel 
functions expression for $u_\pm(z_\pm)$ in the off-diagonal 
elements of (\ref{4.10}), and then notice that the numerator of 
the determinant factorizes into the product 
$(Y_2^-J_2^+-Y_2^+J_2^-)(J_1^+\,Y_1^-- J_1^-\,Y_1^+ )$.} 
    \begin{equation} 
    \det {\bf G}(\Box)=-\frac1{z_+z_-\Box}\, 
    \frac{Y_2^-J_2^+-Y_2^+J_2^-} 
    {Y_1^-J_1^+-Y_1^+J_1^-},         \label{4.10a} 
    \end{equation} 
so that ${\bf F}(\Box)$ assumes a form structurally similar to 
(\ref{4.10}), 
    \begin{equation} 
    {\bf F}(\Box) =-\frac{1}{ J_2^+\,Y_2^- -J_2^-\,Y_2^+} 
    \left[\begin{array}{cc} 
    \,\sqrt{\Box} z_+u_+(z_-)&-\displaystyle{\frac2\pi} \\ 
    -\displaystyle{\frac2\pi}&\sqrt{\Box} z_-u_-(z_+)\, 
    \end{array}\right].                                 \label{4.11} 
    \end{equation} 
This exact expression for ${\bf F}(\Box)$ will be important for us 
in what follows when considering the low-energy approximation, 
because a direct expansion of (\ref{4.11}) in powers of $\Box$ 
turns out to be much simpler than the expansion of (\ref{4.10}) 
with its subsequent inversion. 
 
\subsection{The low-energy limit --- recovery of Einstein theory} 
 
The Green's function (\ref{4.10}) and its inverse operator 
(\ref{4.11}) are nonlinear functions of $\Box$.  This fact corresponds to the 
essentially nonlocal nature of the effective four-dimensional theory induced 
on the branes from the bulk. A remarkable property of the one-brane 
Randall-Sundrum model is however that the low-energy approximation of the 
effective four-dimensional theory is (quasi)local and corresponds to Einstein 
gravity minimally coupled to matter fields on the branes. In the two--brane 
model the situation is more complicated --- the low-energy theory belongs to 
the Brans-Dicke type with a non-minimal curvature coupling of the extra scalar 
field that mediates the interaction of the metric with the trace of the matter 
stress tensor \cite{GT}. 
 
We begin considering the low-energy effective theory in the 
leading order approximation by taking the limit $\Box\to 0$ of the 
Green's function ${\bf G}(\Box)$, (\ref{4.10}).  By using the 
well-known asymptotics of small argument for Bessel and Neumann 
functions one finds the low-energy behavior of various ingredients 
of Eqs.~(\ref{4.10}) and (\ref{4.11}) as expansions in integer and 
half-integer powers of $\Box$ and, thus, obtains the leading order  
approximation in $\Box\to 0$ 
    \begin{equation} 
    {\bf G}(\Box)=\frac2{l^2\Box}\,\frac1{1-a^2} 
    \left[\begin{array}{cc} 
    \,\,1\,\,&\,\,a^2\,\, \\ 
    \,\,a^2\,\, &\,\,a^4\,\, 
    \end{array}\right]+O(l^2\Box),                               \label{4.12} 
    \end{equation} 
    \begin{equation} 
    {\bf G}_\xi(\Box)=-\frac2{l^2\Box}\,\frac{a^2}{1-a^2} 
    \left[\begin{array}{cc} 
    \,\,1\,\,&\,-1\,\, \\ 
    \,\,1\,\,&\,-1\,\, 
    \end{array}\right]+O(l^2\Box),                \label{4.13} 
    \end{equation} 
where $a$ is the brane separation parameter, $a=e^{-d/l}$. 
 
With these Green's functions the effective equations (\ref{2.25}) 
take the form of the following two (linearized) Einstein equations, 
    \begin{eqnarray} 
    &&E_{\mu\nu}^+=8\pi\Big(G^+T_{\mu\nu}^+ 
    +G^-T_{\mu\nu}^-\Big)+\frac1l\, 
    \frac{e^{-d/l}}{\sinh(d/l)}\, 
    (\nabla_\mu\nabla_\nu- 
    \eta_{\mu\nu}\Box)(\xi^+-\xi^-),              \label{4.14a}\\ 
    &&E_{\mu\nu}^-=8\pi e^{-2d/l}\Big(G^+T_{\mu\nu}^+ 
    +G^-T_{\mu\nu}^-\Big)+\frac1l\, 
    \frac{e^{-d/l}}{\sinh(d/l)}\, 
    (\nabla_\mu\nabla_\nu- 
    \eta_{\mu\nu}\Box)(\xi^+-\xi^-).            \label{4.14b} 
    \end{eqnarray} 
Matter sources on both branes are coupled to gravity with the 
effective four-dimensional gravitational constants $G^\pm$ 
\cite{GT} depending on the brane separation, 
    \begin{equation} 
    G^\pm=\frac{G_5}l\, 
    \frac{e^{\pm d/l}}{2\sinh(d/l)},            \label{4.15} 
    \end{equation} 
and there is also a non-minimal curvature coupling to a particular 
combination of radion fields $(\xi^+-\xi^-)$ which, obviously, 
describes dynamical disturbances of the interbrane distance. 
Thus, the low-energy theory reduces to the generalized Brans-Dicke 
model --- the fact that was first observed in \cite{GT} (see also 
\cite{MukKof} on the realization of this property in braneworld 
scenarios with bulk scalar fields). 
 
For large distance between the branes, $a=e^{-d/l}\ll 1$, the metric field 
on the positive-tension brane decouples from all fields on the 
other brane and from radions because 
    \begin{equation} 
    G^+=\frac{G_5}l\, 
    \frac1{1-e^{-2d/l}}\to G_4,\,\,\, 
    G^-\simeq G_4e^{-2d/l}\to 0,\,\,\,d\to\infty,  \label{4.16} 
    \end{equation} 
and the low-energy theory on this brane (usually called Planckian)  
reduces to Einstein 
gravity with the four-dimensional gravitational constant 
(\ref{1.12a}). This is a manifestation of the so-called graviton 
zero-mode localization in the one-brane Randall-Sundrum model 
\cite{RSloc} or the AdS/CFT-correspondence \cite{Gubser,HHR1,HHR2}. 
The recovery of Einstein theory on the Planckian brane is 
the result of a non-trivial cancellation between the contributions 
of the stress tensor trace and the radion in the right-hand side 
of the effective equations of motion (\ref{2.23}). This cancellation 
leads to Eq.~(\ref{2.23a}) with an exponentially small matrix 
${\bf G}_\xi(\Box)\sim a^2\to 0$, (\ref{4.13}), of the non-minimal 
coupling to radions. 
 
For finite interbrane distance, both stress tensors 
$T_{\mu\nu}^\pm$ contribute to the right-hand sides of Einstein's 
equations (\ref{4.14a})--(\ref{4.14b}). They contribute to 
$E_{\mu\nu}^\pm$ with different strengths (their contribution to the 
negative tension brane metric is $e^{-2d/l}$ times weaker), but in 
one and the same combination $G^+T_{\mu\nu}^++G^-T_{\mu\nu}^-$. 
This, maybe physically obvious, fact has a crucial consequence for 
the structure of the off-shell extension of the effective 
braneworld theory. Mathematically this property manifests itself 
in the degeneracy of the leading-order matrix Green's function 
(\ref{4.12}) and results, as we will see below, in the presence of 
a massive graviton mode. Another degeneration that occurs in the 
low-energy limit is the fact that among 
two radion fields, that were introduced on a kinematical ground as 
independent entities, only their combination $(\xi^+-\xi^-)$ is 
dynamical. Apparently, this is the explanation why only one radion 
field is usually considered as a dynamically relevant variable in 
the two-brane Randall-Sundrum model (see \cite{GarPujTan} where 
this property was explained by the homogeneity of the AdS background). 
Such a degeneration, as we see, is not fundamental, but turns out 
to be an artifact of the adopted low-energy approximation scheme. 
 
The braneworld effective action requires the knowledge of ${\bf 
F}(\Box)$ --- the inverse of the Green's function ${\bf G}(\Box)$. 
The latter, as we have just seen, is degenerate in the lowest 
order of the $\Box$-expansion (\ref{4.12}), which is why we have to 
go beyond this approximation. In this expansion one can single out 
two distinctly different physical regimes characterized by the 
values of the dimensionless parameters $l^2\Box$ and $a$. One 
regime corresponds to small energies compared to the AdS-scale, 
$\Box\ll 1/l^2$ or $l^2\Box\ll 1$, and small or finite interbrane 
distance $a=O(1)$, such that $l^2\Box/a^2\ll 1$. The last 
requirement implies that the physical energy range is small also 
at the negative tension brane, because the physical energy at  
this brane is determined by 
$\sqrt{g^{\mu\nu}_-\partial_\mu\partial_\nu}=\sqrt\Box/a$. 
 
Another regime corresponds to small energies on the positive 
tension brane and large energies on the negative tension one, 
$l^2\Box\ll 1$ and $l^2\Box/a^2\gg 1$, when $a\ll 1$. This limit 
includes, in particular, the situation when the second brane is 
pushed to infinity of the fifth coordinate (the AdS horizon) and 
qualitatively is equivalent to the one-brane 
situation\footnote{The situation with an infinitely remote second 
brane is not entirely equivalent to the one-brane model, because 
the Israel junction condition on the second brane is different 
from the usually assumed Hartle boundary conditions on the AdS 
horizon, see Sec.~\ref{hartle}.}. 
 
\subsection{Low-energy derivative expansion  
\label{particle}} 
Here we consider the derivative expansion of the nonlocal form factors 
in the first of the low-energy regimes, corresponding to 
small or finite interbrane distance. In this regime the arguments 
of both sets of Bessel and Neumann functions in 
(\ref{4.10})--(\ref{4.11}) are small 
    \begin{equation} 
    l\sqrt\Box\ll 1,\,\,\,\, \frac{l\sqrt\Box}a\ll 1.  \label{4.19} 
    \end{equation} 
Therefore, one can expand ${\bf G}(\Box)$ to higher than 
zeroth order in $\Box$ and, thus, make it explicitly invertible. 
Since ${\bf G}(\Box)$ is degenerate in the zeroth order, its 
matrix determinant is at least $O(1/\Box)$ rather than 
$O(1/\Box^2)$ and, therefore, one should expect that its inverse 
${\bf F}(\Box)$ will be a massive operator 
--- its expansion in powers of $\Box$ will start with the mass 
matrix $O(\Box^0)$. This degeneracy leads also to an additional 
difficulty --- in order to achieve the kinetic term in ${\bf 
F}(\Box)$ linear in $\Box$ one would have to calculate the Green's 
function ${\bf G}(\Box)$ to $O(\Box^2)$ inclusive.  
Fortunately, instead of inverting the Green's function expansion, 
we have the exact expression (\ref{4.11}) for ${\bf F}(\Box)$ that 
can be directly expanded to a needed order. Thus, using the small  
argument expansions of Bessel functions in (\ref{4.11}) we get 
    \begin{eqnarray} 
    &&\frac{\bf F(\Box)}{l^2}= 
    -{\bf M}_F+{\bf D}_F\Box+ 
    {\bf F}^{(2)}l^2\Box^2+O(\Box^3),    \label{4.26}\\ 
    &&{\bf M}_F=\frac1{l^2}\,\frac4{1-a^4}\, 
    \left[\begin{array}{cc} 
    \,a^4\, & \,-a^2 \\ 
    -a^2 & \,1 \end{array}\right],      \label{4.27}\\ 
    &&{\bf D}_F=\frac{1-a^2}{6(1+a^2)^2} 
    \left[\begin{array}{cc} 
    \,a^2+3\, & 2 \\ 
    2 & \,3+a^{-2} 
    \end{array}\right]\ ,                 \label{4.28} 
    \end{eqnarray} 
where the components of the matrix ${\bf F}^{(2)}$ are rather  
lengthy (but we will need them below because they will  
qualitatively effect the low-energy behaviour in the radion  
sector), 
    \begin{eqnarray} 
    &&F^{(2)}_{++} = 
    \frac{(1-a^2)^3(3+a^2)}{72a^2(1+a^2)^3} 
    -\frac{\ln a}{4(1-a^4)^2}- 
    \frac{4+5a^2-4a^4+a^6}{96a^2(1-a^4)},\nonumber\\ 
    &&F^{(2)}_{--} = 
    \frac{(1-a^2)^3(3a^2+1)}{72a^4(1+a^2)^3} 
    -\frac{a^4\ln a}{4(1-a^4)^2} 
    -\frac{1-4a^2+5a^4+4a^6}{96a^4(1-a^4)},\nonumber\\ 
    &&F^{(2)}_{+-} = 
    \frac{(1-a^2)^3}{36a^2(1+a^2)^3} 
    +\frac{a^2\ln a}{4(1-a^4)^2} 
    -\frac{1-8a^2+a^4}{96a^2(1-a^4)}.   \label{4.21} 
    \end{eqnarray} 
The matrix coefficients of this $\Box$-expansion are non-diagonal 
and, therefore, nontrivially entangle the fields on both branes.  
An important property of the lowest-order coefficient --- the mass  
matrix ${\bf M}_F$ --- is that it is degenerate and has rank one. As  
we will see below, this fact will guarantee the presence of one  
massless graviton in the spectrum of the braneworld action. 
 
The low-energy expansion for ${\bf K}(\Box)$ follows from that of  
the operator ${\bf F}(\Box)$, (\ref{4.26})--(\ref{4.21}), 
    \begin{eqnarray} 
    &&\frac{{\bf K}(\Box)}{l^2}=-{\bf M}_K+ 
    {\bf D}_K \Box 
    +{\bf K}^{(2)}\,l^2\Box^2+O(\Box^3)\ ,  \label{7.3.3}\\ 
    &&{\bf M}_K=2{\bf M}_F= 
    \frac8{l^2}\frac1{1-a^4} 
    \left[\begin{array}{cc} 
    \,\,a^4\, & \,-a^2 \\ 
    \,-a^2 & \,\,1\,\end{array}\right],   \label{7.3.4}\\ 
    &&{\bf D}_K=\frac2{3(1+a^2)^2} 
    \left[\begin{array}{cc} 
    \,-4a^2-2a^4\, & 1-a^2 \\ 
    1-a^2 & \,4+2a^{-2}\, 
    \end{array}\right],                 \label{4.22} 
    \end{eqnarray} 
where ${\bf K}^{(2)}=2{\bf F}^{(2)}$. Similarly to the graviton 
operator ${\bf F}(\Box)$, the mass matrix ${\bf M}_K$ is 
degenerate. Moreover, the matrix determinant of the radion operator 
is at least quadratic in $\Box$ --- the property  
responsible for the dipole-ghost nature of the radion field (see  
Sect.\ref{reduced}), 
    \begin{eqnarray} 
    \det{\bf K}(\Box)= 
    \frac{4\ln a}{1-a^4}\,(l^2\Box)^2 
    +O[\,(l^2\Box)^3].                \label{7.3.5} 
    \end{eqnarray}

\subsection{Particle spectrum of the braneworld 
action and Kaluza-Klein modes \label{massless}} 
The quadratic approximation for the action and its nonlocal formfactors 
obviously determines the spectrum of excitations in the theory. Here we 
show that in the graviton sector this spectrum corresponds to the 
tower of Kaluza-Klein modes well-known from a conventional Kaluza-Klein 
setup. The graviton sector arises when one decomposes metric perturbations  
on both branes into irreducible components --- transverse-traceless  
tensor, vector and scalar parts, 
    \begin{equation} 
    h_{\mu\nu}^\pm=\gamma_{\mu\nu}^\pm 
    +\varphi^\pm\eta_{\mu\nu} 
    +f_{\mu,\nu}+f_{\nu,\mu}\,,\,\,\, 
    {\gamma_{\mu\nu}}^{,\,\nu} 
    =\eta^{\mu\nu}\gamma_{\mu\nu}=0.    \label{4.25} 
    \end{equation} 
On substituting this decomposition in the linearized curvatures of  
(\ref{1.12}) one finds that the vector parts do not contribute  
to the action, and the latter reduces to the sum of the graviton and 
scalar sectors, 
    \begin{equation} 
    S_4\,[\,g^\pm_{\mu\nu},\psi^\pm]= 
    S_{\rm graviton}[\gamma^\pm_{\mu\nu}]+ 
    S_{\rm scalar}[\,\varphi^\pm,\psi^\pm]. 
    \end{equation} 
The graviton part is entirely determined by the operator ${\bf 
F}(\Box)$ and reads 
    \begin{equation} 
    S_{\rm graviton}[\gamma^\pm_{\mu\nu}] 
    =\frac1{16\pi G_4}\int 
    d^4x\,\frac12\, 
    [\,\gamma_{\mu\nu}^+\,\,\,\gamma_{\mu\nu}^-\,] 
    \,\frac{\bf F(\Box)}{l^2}\, 
    \left[\begin{array}{c} 
    \,\,\gamma^{\mu\nu}_+ \\ 
    \,\,\gamma^{\mu\nu}_-\, 
    \end{array}\right],        \label{4.24} 
    \end{equation} 
while the scalar sector consists of the radion fields of Eq.~(\ref{1.13a}) 
and the doublets of the trace (or conformal) parts of the metric 
perturbations $\varphi^\pm$, 
    \begin{equation} 
    {\bf\Phi}=\left[\begin{array}{c} 
    \,\,\varphi^+(x)\,\\ 
    \,\,\varphi^-(x)\, 
    \end{array}\right],\,\,\, 
    {\bf\Phi}^T=\Big[\,\varphi^+(x)\,\,\, 
    \varphi^-(x)\,\Big]. 
    \end{equation} 
Their action diagonalizes in terms of  
the conformal modes and the (redefined) radion modes  
$2{\bf\Psi-\Phi}$, 
    \begin{eqnarray} 
    &&S_{\rm scalar}[\,\varphi^\pm,\psi^\pm] 
    =\frac3{32\pi G_4} \int d^4x\, 
    \Big(-\varphi^+\Box\varphi^+ 
    +\frac1{a^2}\,\varphi^-\Box\varphi^-\Big)\nonumber\\ 
    &&\qquad\qquad\qquad\qquad\qquad\quad 
    -\frac3{16\pi G_4} \int d^4x\, 
    \big(\,2{\bf\Psi}-{\bf\Phi}\big)^T 
    \frac{{\bf K}(\Box)}{l^2} 
    \big(\,2{\bf\Psi}-{\bf\Phi}\big).       \label{4.24a} 
    \end{eqnarray} 
Note that the sector of conformal modes is entirely local, and the 
last $2\times2$ radion quadratic form here corresponds to the last term 
of (\ref{1.12}) quadratic in the left-hand side of the radion equations of  
motion (\ref{2.29}), 
$\Box{\bf\Psi}+{\bf R}/6=\Box(2{\bf\Psi-\Phi})/2$. 
 
Excitations in the graviton sector are the transverse-traceless 
basis functions ${\bf v}(x)={\bf v}_{\mu\nu}(x)$ of the operator  
${\bf F}(\Box)$, 
    \begin{equation} 
    {\bf F}(\Box)\,{\bf v}_n(x)=0,\,\,\,\,\, 
    {\bf v}_n(x)=\left[\begin{array}{c} 
    v^+_n(x)\\ 
    v^-_n(x) 
    \end{array}\right].            \label{4.32b} 
    \end{equation} 
The existence condition for zero-vectors of the $2\times2$  
matrix operator ${\bf F}(\Box)$,  
    \begin{equation} 
    {\rm det}\,{\bf F}(\Box)=0,    \label{4.32} 
    \end{equation} 
serves as the equation for the masses of these propagating modes  
$m_n$, 
    \begin{equation} 
    (\Box-m_n^2)\,{\bf v}_n(x)=0,    \label{4.32a} 
    \end{equation} 
determined by the roots $\Box=m_n^2$ of (\ref{4.32}). In view of  
Eq.~(\ref{4.10a}) these roots coincide with the zeros  
of the following combination of Bessel functions 
    \begin{equation} 
    \Box\,\Big(Y_1^-J_1^+-Y_1^+J_1^-\Big)=0.  \label{4.33} 
    \end{equation} 
But the same equation (\ref{4.33}) determines the 
spectrum of the Kaluza-Klein modes --- the zeros of the Wronskian  
of basis functions $u_\pm(z)$, Eq.~(\ref{4.8}), entering the Neumann 
Green's function (\ref{4.7}). Therefore, a conventional tower of  
massive Kaluza-Klein modes is contained in the spectrum of the braneworld 
effective action. 
 
The derivative expansion of the previous section allows one to  
get a reliable description for the  
massless sector of the spectrum, $m_0^2=0$. In this sector 
Eq.~(\ref{4.32b}) reduces to ${\bf M}_F{\bf v}_0(x)=0$ and implies  
that ${\bf v}_0(x)$ is a zero mode of the mass matrix ${\bf M}_F$ ---  
the property guaranteed by the degeneracy of this matrix, mentioned  
above. The non-diagonal nature of (\ref{4.27}) implies that 
this mode is still the collective excitation of tensor perturbations 
$v^\pm_0(x)$ on both branes, but $v^-_0(x)=a^2 v^+_0(x)<v^+_0(x)$ and 
for large interbrane separation, $a\to 0$, the negative brane component 
tends to zero, $v^-_0(x)\to 0$ , so that the massless graviton is 
essentially localized on the positive-tension brane. This is, certainly, 
another manifestation of the recovery of Einstein theory on this brane 
for the two-brane Randall-Sundrum model. 
 
The attempt to describe massive modes within the derivative expansion 
for ${\bf F}(\Box)$ turns out to be inconsistent. Indeed, the truncation 
of the series (\ref{4.26}) on the second term, for the goal of finding 
the first massive level, leads to the equation 
${\rm det}(-{\bf M}_F+{\bf D}_F\Box)=0$ instead of (\ref{4.32}). It 
implies the simultaneous diagonalization of both mass and kinetic term 
matrices in the basis of the first two (massless and massive) propagating  
modes (which is possible in view of the positive-definiteness 
of the kinetic matrix ${\bf D}_F$). But, unfortunately, the massive  
root of this equation,  
$\Box=M^2\equiv 24a^2(1+a^2)/l^2(1-a^2)^2\gg a^2/l^2$, strongly violates  
the second of the low-energy conditions (\ref{4.19}). The excited 
massive graviton mode always turns out to be in the physical high-energy 
domain on the negative-tension brane, and the low-energy description 
fails. For this reason in this paper we restrict ourselves to the 
massless sector of the theory --- the effects of massive modes will be  
considered elsewhere \cite{rigo}.

\subsection{Large interbrane distance\label{large}} 
In view of the discussion of the previous section, it is instructive  
to consider in the low-energy approximation on the positive-tension brane 
the case of large brane separation, when $a\ll 1$ and 
    \begin{equation} 
    l\sqrt\Box\ll 1,\,\,\, \frac{l\sqrt\Box}a\gg 1. \label{7.4.1} 
    \end{equation} 
This range of coordinate distances $1/\sqrt\Box$ corresponds to the  
long-distance approximation on the $\Sigma_+$-brane and to the physical  
{\em short}-distance limit, $a/\sqrt\Box\ll l$, on the $\Sigma_-$-brane. 
Now one should use the asymptotic expressions of large arguments of 
the Bessel functions $(J^-_\nu,Y^-_\nu)$, $\nu=1,2$, 
    \begin{eqnarray} 
    J^-_\nu\simeq\sqrt{\frac{2a}{\pi l\Box^{1/2}}} 
    \cos\left(\frac{l\sqrt\Box}a-\frac\pi4- 
    \frac{\pi\nu}2\right),\,\,\, 
    Y^-_\nu\simeq\sqrt{\frac{2a}{\pi l\Box^{1/2}}} 
    \sin\left(\frac{l\sqrt\Box}a-\frac\pi4- 
    \frac{\pi\nu}2\right)\ ,                      \label{7.4.2} 
    \end{eqnarray} 
and as before the small-argument expansions for $(J^+_\nu,Y^+_\nu)$. 
Then, in the leading order the operator ${\bf F}(\Box)$ reads 
    \begin{equation} 
    {\bf F}(\Box) \simeq 
    \left[\begin{array}{cr} 
    \,\,\displaystyle{l^2\Box\over2}\,\,  
    & \,\,\displaystyle{l^2\Box\over 2J_2^-}\,\,\,\,\, \\ 
    & \\ 
    \,\,\,\displaystyle{l^2\Box\over 2J_2^-}\,\, & 
    \,\, -\displaystyle{J_1^-\over J_2^-}\, 
    \frac{l}{a}\sqrt{\Box}\end{array}\right].      \label{7.4.3} 
    \end{equation} 
In contrast to the case of small brane separation, the 
short-distance corrections to this matrix operator contain 
a nonlocal $\Box^2\ln\Box$-term. Here we present it for the 
$F_{++}(\Box)$-element, 
    \begin{eqnarray} 
    &&F_{++}(\Box)=\frac{l^2\Box}2 
    +\frac{(l^2\Box)^2}2\,k_2(\Box) 
    +O\Big[\,(l^2\Box)^3\Big],                   \label{7.4.4}\\ 
    &&k_2(\Box)=\frac14\left(\,\ln\frac4{l^2\Box} 
    -2{\bf C}+ 
    \pi\frac{Y_2^-}{J_2^-}\,\right)          \label{7.4.5} 
    \end{eqnarray} 
(the meaning of the subscript in $k_2(\Box)$ will become 
clear below). This is a manifestation of the well-known phenomenon 
of AdS/CFT-correspondence \cite{AdS/CFT,Gubser,logcoef} when 
typical quantum field theoretical effects in four-dimensional 
theory can be generated from the classical theory in the bulk. The 
AdS/CFT-duality exists for the boundary theory in the AdS bulk when 
the brane, treated as a boundary of the AdS spacetime, tends to 
infinity. In the wording of the two-brane Randall-Sundrum 
model \cite{Gubser,brane} this situation qualitatively corresponds 
to the case of the negative tension brane tending to the horizon 
of the AdS spacetime, $y\to\infty$, or $a\ll 1$. 
 
The non-logarithmic term of Eq.~(\ref{7.4.5}) include the 
functions with an infinite series of poles at  
$l^2\Box\simeq\pi^2 a^2(n+3/4)^2$, $n=0,1,...$, contained in the ratio 
    \begin{eqnarray} 
    \frac{Y_2^-}{J_2^-}\simeq 
    -\frac{J_1^-}{J_2^-}\simeq 
    \tan\left(\frac{l\sqrt\Box}a 
    -\frac\pi4\right).              \label{7.4.6} 
    \end{eqnarray} 
Interestingly, these poles arise in the wave operator of the 
theory, rather than only in its Green's function. This is an 
artifact of the nonlocality, when both the propagator and its inverse 
are nonlocal. The poles are separated by the sequence of roots on the  
positive real axis of the $\Box$-plane,  
$\Box=m_n^2\simeq(\pi^2a^2/l^2)(n+3/4)^2/l^2\big(1 
+\pi^2a^2(n+3/4)/2+...\big)$, 
which correspond to the tower of massive Kaluza-Klein modes in the  
energy range (\ref{7.4.1})\footnote{In this range the above expressions 
for poles and zeros of ${\bf F}(\Box)$ are valid for large $n$ satisfying 
$1\ll\pi(n+3/4)\ll 1/a$.}. They are excited for large brane separation,  
$a\ll 1$, because a small energy range at the positive tension brane  
turns out to be the high energy range at the second brane.  
 
The presence 
of both zeros and poles in the wave operator ${\bf F}(\Box)$ can, 
apparently, be explained by the duality relation between the Dirichlet 
and Neumann Green's functions in braneworld physics \cite{duality}.  
According to this relation the Neumann Green's function in the bulk  
when restricted to branes (which is exactly ${\bf F}(\Box)$) is the  
inverse of the bulk Dirichlet Green's function properly differentiated  
with respect to its two arguments and also restricted to branes. Therefore, 
the necessarily existing poles of the Dirichlet Green's function generate  
zeros of the Neumann one and vicy versa. The presence of zeros, each of 
which being located between a relevant pair of neighbouring poles of  
${\bf F}(\Box)$, is actually very important, because this guarantees the 
positivity of residues of all poles in ${\bf G}(\Box)={\bf F}^{-1}(\Box)$ 
or the normal non-ghost nature of all massive modes.

\section{The reduced effective action\label{reduced}} 
If we take the usual viewpoint of the braneworld 
framework, that our visible world is one of the branes embedded in 
a higher-dimensional bulk, then the fields living on other branes 
are not directly observable. In this case the effective dynamics should be 
formulated in terms of fields on the visible brane. In the two-brane 
Randall-Sundrum model this is equivalent to constructing the reduced 
action --- an action with on-shell reduction for the invisible fields 
in terms of the visible ones. This reduction is nothing but the tree-level 
procedure of tracing (or integrating) out the unobservable variables. 
It implies that in the two-brane action we have to exclude the  
fields on the invisible brane in 
terms of those on the visible one. As the latter we choose the 
positive-tension brane. One of the reasons for such a choice is 
that the low-energy dynamics on this brane is closest to 
four-dimensional Einstein theory, while that of the 
negative-tension brane is encumbered with the problems discussed 
above --- intrusion into the high-energy domain, excitation of massive 
modes, etc. 
 
We perform the reduction of the action to the $\Sigma_+$-fields 
separately in the graviton and scalar sectors. In the graviton 
sector (\ref{4.24}) the on-shell reduction --- the exclusion of  
$\gamma^-_{\mu\nu}$-perturbations in terms of $\gamma_{\mu\nu}^+ 
=\gamma_{\mu\nu}$ (in what follows we omit the label $+$, because  
only one field remains) --- corresponds to the replacement of the  
original action by the new one, 
    \begin{eqnarray} 
    S_{\rm graviton}[\,\gamma^\pm_{\mu\nu}]\,\Rightarrow\, 
    S_{\rm graviton}^{\rm red}[\,g_{\mu\nu}] 
    =\frac1{16\pi G_4}\int 
    d^4x\,\sqrt g 
    \gamma_{\mu\nu}^+ 
    \frac{F_{\rm red}(\Box)}{2\,l^2} 
    \gamma^{\mu\nu}_+ ,                     \label{8.1} 
    \end{eqnarray} 
with the original kernel ${\bf F}(\Box)$ going over to the new  
one-component kernel $F_{\rm red}(\Box)$ according to the  
following simple prescription 
    \begin{eqnarray} 
    {\bf F}(\Box)\, \Rightarrow\, 
    F_{\rm red}(\Box)=F_{++}(\Box)- 
    F_{+-}(\Box)\frac1{F_{--}(\Box)}F_{-+}(\Box). \label{8.2} 
    \end{eqnarray} 
It is useful to rewrite (\ref{8.1}) back to the covariant form in 
terms of (linearized) Ricci curvatures on a single visible brane, 
    \begin{equation} 
    S_{\rm graviton}^{\rm red}[g_{\mu\nu}] 
    =\frac1{8\pi G_4}\int 
    d^4x\,\sqrt g\left(R_{\mu\nu} 
    \frac{F_{\rm red}(\Box)}{l^2\,\Box^2}\, 
    R^{\mu\nu} 
    -\frac13\,R\frac{F_{\rm red} 
    (\Box)}{l^2\,\Box^2}\,R\right)\ .           \label{8.3} 
    \end{equation} 
 
A similar reduction in the scalar sector implies omitting in 
the first integral of (\ref{4.24a}) the negative-tension term  
and replacing the $2\times2$ quadratic form in the second integral  
by the quadratic form in $\psi^+$ with the reduced operator 
    \begin{eqnarray} 
    K_{\rm red}(\Box)=K_{++}(\Box)- 
    K_{+-}(\Box)\frac1{K_{--}(\Box)}K_{-+}(\Box) 
    =\frac{\det{\bf K}(\Box)}{K_{--}(\Box)}\ .    \label{8.8} 
    \end{eqnarray} 
 
Finally, we express the conformal mode in terms of the 
(linearized) Ricci scalar $\varphi^+=-(1/3\Box)R$, and denote the 
radion by $\psi^+=\psi$. Then the combination of the reduced scalar 
sector together with the graviton part (\ref{8.3}) yields the reduced 
action in its covariant form 
    \begin{eqnarray} 
    &&S_{\rm red}[g_{\mu\nu},\psi]=\frac1{16\pi G_4} 
    \int d^4x\,\sqrt{g}\left[\,R_{\mu\nu} 
    \frac{2F_{\rm red}}{l^2\Box^2} 
    \left(R^{\mu\nu}-\frac12 g^{\mu\nu}R\right) 
    \right.\nonumber\\ 
    &&\qquad\qquad\qquad\quad\left. 
    -\frac16R\left(\frac1\Box 
    -\frac{2F_{\rm red}}{l^2\Box^2}\right)R 
    -6l^2\left(\Box\psi+\frac R6\right) 
    \frac{2K_{\rm red}}{l^2\Box^2} 
    \left(\Box\psi+\frac R6\right)\,\right].  \label{8.9} 
    \end{eqnarray} 
Here we have deliberately singled out the term bilinear in the Ricci 
and Einstein tensors, because this form will be useful in 
comparing the low-energy approximation of this action with the 
Einstein action. Below we analyze this result in the two low-energy 
regimes (\ref{4.19}) and (\ref{7.4.1}). 
 
\subsection{Small interbrane distance} 
In the regime of small or finite brane 
separation (\ref{4.19}), the calculation of the reduced operator 
(\ref{8.2}) gives, on using (\ref{4.26})--(\ref{4.28}), a very simple 
result, 
    \begin{eqnarray} 
    &&F_{\rm red}(\Box)=\frac{l^2\Box}2(1-a^2) 
    +\frac{(l^2\Box)^2}2\kappa_1(a) 
    +O[\,(l^2\Box)^3],                   \label{8.10}\\ 
    &&\kappa_1(a)=\frac14\, 
    \left[\,\ln\frac1{a^2}-(1-a^2)- 
      \frac12 (1-a^2)^2\,\right].       \label{8.10a} 
    \end{eqnarray} 
The reduced operator turns out to be massless, and this is a 
corollary of the degenerate nature of the mass matrix 
(\ref{4.27}), $\det{\bf M}_F=0$, because  
$F_{\rm red}(\Box)=\det{\bf F}(\Box)/F_{--}(\Box)=O(\Box)$.  
Similary, in view of (\ref{7.3.5}), the reduced operator in  
the radion sector (\ref{8.8}) is at least quadratic in $\Box$, 
    \begin{eqnarray} 
    &&K_{\rm red}(\Box)= 
    \kappa_2(a)(l^2\Box)^2 
    +O[\,(l^2\Box)^3],             \label{8.14}\\ 
    &&\kappa_2(a)= 
    \frac14\ln\frac1{a^2},           \label{8.15} 
    \end{eqnarray} 
so that the low-energy radion turns out to be a dipole ghost.

After substituting (\ref{8.10}) 
into (\ref{8.9}), the first term bilinear in $R_{\mu\nu}$ and 
$R^{\mu\nu}-\frac12g^{\mu\nu}R$ seems to remain nonlocal. However, 
this term is nothing but the part of the local Einstein action, 
which is quadratic in metric perturbations $\delta 
g_{\mu\nu}=h_{\mu\nu}$ on a flat spacetime background. To see 
this, note that, up to diffeomorphism with some vector field 
parameter $f_\mu$, this perturbation can be nonlocally rewritten 
in terms of the (linearized) Ricci tensor, 
    $h_{\mu\nu}=-(2/\Box)R_{\mu\nu}+ 
    f_{\mu\,,\,\nu}+ 
    f_{\nu\,,\,\mu}+O[\,R^2_{\mu\nu}\,]$. 
When substituted into the quadratic part of the Einstein action it 
takes an explicitly nonlocal form in terms of Ricci curvatures, 
    \begin{equation} 
    \frac12\,\delta^2\!\!\int d^4x\, 
    \sqrt{g}R=\int d^4x\,\sqrt{g} 
    \left(R^{\mu\nu}-\frac12 
    g^{\mu\nu}R\right)\frac1\Box R_{\mu\nu} 
    +O[\,R^3_{\mu\nu}\,],                  \label{8.12} 
    \end{equation} 
which is exactly the first term of (\ref{8.9}) with the first-order 
in $\Box$ approximation for $F_{\rm red}(\Box)$, (\ref{8.10}). The part 
of the Einstein action linear in perturbations is a total 
divergence, which we disregard here, and the zeroth order term is 
identically vanishing. Therefore, this term can be rewritten as 
the local Einstein action linear in scalar curvature with the 
$a$-dependent four-dimensional gravitational constant 
(\ref{4.16}), $G_4(a)=G_4/(1-a^2)=G^+$. The second term of 
(\ref{8.9}) with $F_{\rm red}(\Box)$ taken in the same approximation  
stays nonlocal although this nonlocality is suppressed by the factor 
$a^2<1$. 
 
The quadratic in $\Box$ contribution to (\ref{8.10}) generates the  
local part of the action quadratic in curvatures which enter in  
a special combination $R^2_{\mu\nu}-R^2/3$. This combination 
differs from the square of the Weyl tensor $C_{\mu\nu\alpha\beta}^2$ 
by the density of the Gauss-Bonnet invariant  
$E=R_{\mu\nu\alpha\beta}^2-4R^2_{\mu\nu}+R^2$, 
    $2(R^2_{\mu\nu}-R^2/3) 
    =C_{\mu\nu\alpha\beta}^2-E$, 
which can be omitted under the integral sign (because it is  
topologically invariant and in the quadratic order in metric 
perturbations explicitly reduces to the total surface term  
\cite{CPT,CPTIII}). 
 
Thus, collecting different contributions together, we get 
the reduced braneworld action in the low-energy regime of finite 
interbrane distance 
    \begin{eqnarray} 
    &&S_{\rm red}[\,g_{\mu\nu},\psi\,]=\frac1{16\pi G_4} 
    \int d^4x\,\sqrt{g} 
    \left[\,(1-a^2)R 
      -\frac{a^2}6\,R\frac1\Box R\right.\nonumber\\ 
    &&\qquad\qquad\qquad\qquad\qquad 
     \left. -6l^2\kappa(a) 
    \Big(\Box\psi+\frac R6\Big)^2 
    +\frac{l^2}2\,\kappa_1(a)\, 
    C_{\mu\nu\alpha\beta}^2\,\right].  \label{8.16} 
    \end{eqnarray} 
The first term here confirms the recovery of Einstein theory on 
the positive tension brane with the well-known expression for the 
effective gravitational constant $G_4(a)$ \cite{GT,ChGR}. The  
higher derivative nature of the 
radion does not really imply physical instability, because $\psi$  
can hardly be treated as non-gauge variable\footnote{Indeed, in  
Sec.~\ref{eom} radions were introduced as gauge variables relating the 
Randall-Sundrum coordinate system to two Gaussian systems associated 
with two branes. Then it was demanded that the kinematical relations 
between radions and stress tensor traces (\ref{2.15a}) should be  
generated as dynamical equations from the braneworld action (\ref{1.12}).  
This has led to the last term in (\ref{1.12}) quadratic in  
$(\Box{\bf\Psi}+{\bf R}/6)$. Thus, this term can be regarded as the  
result of the off-shell extension in the radion sector. The on-shell  
reduction simply corresponds to the exclusion of radions in terms of  
the metric fields or, equivalently, to omitting the last term in  
(\ref{1.12}).}. Its equation of motion, 
    $\Box\left(\Box\psi+R/6\right)=0$, 
implies that the on-shell restriction of (\ref{8.16}) leaves us with 
the first two metric-field terms.  
 
In the low-energy regime the first three terms dominate over 
the local short-distance Weyl-squared part. They form the action that 
was derived in \cite{brane} by a simplified (and, strictly 
speaking, not very legitimate) method --- by just freezing to zero 
all field perturbations on the invisible brane. Here this 
derivation is justified within a consistent scheme accounting for 
the fact that, even without matter sources on $\Sigma_-$, the 
field perturbations on the visible brane induce nontrivial fields 
on the invisible one, and they contribute to the full effective 
action. Interestingly, however, the result turns out to be 
the same as in \cite{brane}. 
 
The second term of (\ref{8.16}) is nonlocal. However, according to 
the discussion in \cite{brane}, it can be localized in terms of an extra 
scalar field. Actually, this extra field can be identified with 
the radion itself up to some nonlocal reparametrization. Indeed, 
the following reparametrization from $\psi$ to the new field $\varphi$, 
    \begin{equation} 
    \varphi=\sqrt{\frac3{4\pi G_4}}\,\left[\,a\Big(1 
      -\frac1{6\Box}R\Big)- 
    l\,\sqrt{\frac{\kappa_2(a)}{-\Box}}\, 
    \Big(\,\Box\psi+\frac16 R\,\Big)\,\right]       \label{8.18} 
    \end{equation} 
converts the action (\ref{8.16}) to the local form  
    \begin{eqnarray} 
    S_{\rm red}[\,g_{\mu\nu},\varphi\,]= 
    \int d^4x\,\sqrt{g} 
    \left[\,\left(\frac1{16\pi G_4} 
     -\frac1{12} \varphi^2\right)R 
     +\frac12\varphi\Box\varphi 
     +\frac{l^2}{32\pi G_4}\,\kappa_1(a)\, 
    C_{\mu\nu\alpha\beta}^2\,\right]       \label{8.19} 
    \end{eqnarray} 
whose first three (low-derivative) terms were derived in  
\cite{brane} by a simplified procedure. 
The field $\varphi$ introduced here by the formal transformation 
(\ref{8.18}) directly arose in \cite{brane} as a local redefinition  
of the radion field relating the Randall-Sundrum coordinates to the 
Gaussian normal coordinates associated with the positive tension 
brane. It is non-minimally coupled to the curvature, and in 
\cite{brane} it was used to play the role of the inflaton 
generating inflation in the presence of a small detuning between the 
values of the brane tensions $\sigma_\pm$ from their Randall-Sundrum 
values (\ref{1.1}). Initial conditions for inflation in 
\cite{brane} were suggested within the tunneling wavefunction 
scheme \cite{tun} modified according to the braneworld creation 
framework \cite{GarSas,HHR1,HHR2}. In this framework the 
Lorentzian spacetime arises as a result of analytic continuation 
from the Euclidean space describing the classically forbidden 
(underbarrier) state of the gravitational field. Note, in 
connection with this, that the transformation (\ref{8.18}) is 
well-defined only in Euclidean spacetime with the 
negative-definite operator $\Box$. Thus, the justification of the 
off-shell reparametrization between the actions (\ref{8.16}) and 
(\ref{8.19}) comes from the Euclidean version of the theory, which 
underlies the braneworld creation scheme\footnote{The 
transformation (\ref{8.18}) is complex-valued for timelike 
momenta, but its on-shell restriction in the radion sector is 
real.}. In the next section we shall extend this justification 
even further by resorting to Hartle boundary conditions on the AdS 
horizon. 
 
\subsection{Large interbrane distance and Hartle boundary conditions 
\label{hartle}} 
 
In the limit $a\to 0$ the nonlocal and correspondingly 
non-minimal terms of (\ref{8.16}) and (\ref{8.19}) vanish and the low-energy 
model seems to reproduce the Einstein theory. However, this limit corresponds 
to another energy regime (\ref{7.4.1}) in which one should use the expressions 
(\ref{7.4.3})--(\ref{7.4.5}) in order to obtain the reduced operator 
(\ref{8.2}).  Then the latter, up to quadratic in $\Box$ terms inclusive, 
reads as 
    \begin{equation} 
    F_{\rm red}(\Box)=\frac{l^2\Box}2 
    +\frac{(l^2\Box)^2}8\, 
    \left(\,\ln\frac4{l^2\Box}-2{\bf C}\,\right) 
    +\frac{(l^2\Box)^2}2 
    \left(\frac\pi4\frac{Y_2^-}{J_2^-}+ 
    \frac{a}{2l\sqrt{\Box}J_1^-J_2^-}\right).  \label{8.20} 
    \end{equation} 
As in (\ref{7.4.4}) it involves the logarithmic nonlocality 
(\ref{7.4.5}) in $\Box^2$-terms. Moreover, the last term here 
simplifies to the ratio of the first order Bessel functions 
$Y_1^-/J_1^-$, so that $F_{\rm red}(\Box)$ takes a form very 
similar to that of the large interbrane separation (\ref{7.4.4}), 
    \begin{eqnarray} 
    F_{\rm red}(\Box)=\frac{l^2\Box}2 
    +\frac{(l^2\Box)^2}8\left[\, 
    \ln\frac4{l^2\Box}-2{\bf C} 
    +\pi\frac{Y_1^-}{J_1^-}\,\right] 
    +O\Big[\,(l^2\Box)^3\Big]\ .              \label{8.21} 
    \end{eqnarray} 
The calculation of the radion operator (\ref{8.8}) with ${\bf 
K}(\Box)$ following from (\ref{7.4.3}) for $l^2\Box/a^2\gg1$ 
results in 
    \begin{eqnarray} 
    K_{\rm red}(\Box)=(l^2\Box)^2\,k_2(\Box),  \label{8.22} 
    \end{eqnarray} 
where $k_2(\Box)$ is defined by (\ref{7.4.5}). 
 
Thus, $F_{\rm red}(\Box)$ and $K_{\rm red}(\Box)$ are given by the 
following two nonlocal operators, 
    \begin{equation} 
    k_\nu(\Box)=\frac14\left[\,\ln\frac4{l^2\Box} 
    -2{\bf C}+ 
    \pi\frac{Y_\nu^-}{J_\nu^-}\,\right], 
    \,\,\,\,\,\nu=1,2,                            \label{8.23} 
    \end{equation} 
and the reduced (one-brane) action finally reads 
    \begin{eqnarray} 
    &&S_{\rm red}[g_{\mu\nu},\psi] 
    =\frac1{16\pi G_4}\int d^4x\,\sqrt{g} 
    \left[\,R+\frac{l^2}2\, 
      C_{\mu\nu\alpha\beta}\,k_1(\Box) 
      C^{\mu\nu\alpha\beta} 
      \right. \nonumber\\ 
    &&\qquad\qquad\qquad\qquad\qquad\qquad\qquad\qquad 
    \left.-6l^2\left(\Box\psi+ 
    \frac R6\right)k_2(\Box) 
    \left(\Box\psi+\frac R6\right)\,\right]. \label{8.24} 
    \end{eqnarray} 
 
Here terms quadratic in curvature (which we again rewrote in terms 
of the Weyl squared combination in view of integration by parts) 
represent short distance 
corrections with form factors whose logarithmic parts have 
an interpretation in terms of the AdS/CFT-correspondence 
\cite{AdS/CFT,Gubser,logcoef,brane}. Their Bessel-function parts 
are more subtle for interpretation and less universal. When taken 
literally they give rise to the massive resonances discussed above in 
Sec.~\ref{large}. However, with the usual Wick rotation prescription 
$\Box\to\Box+i\varepsilon$ these ratios tend to 
    \begin{eqnarray} 
    \frac{Y_\nu^-}{J_\nu^-}\simeq 
    \tan\left(\frac{l}a\sqrt{\Box+i\varepsilon} 
    -\frac\pi4-\frac{\pi\nu}2\right)\to i, 
    \,\,\,a\to 0,                   \label{8.25} 
    \end{eqnarray} 
and both form factors (\ref{8.23}) for $\Box<0$ (Euclidean or 
spacelike momenta) become real and can be expressed in terms of one 
Euclidean form factor as 
    \begin{eqnarray} 
    k_\nu(\Box+i\varepsilon)\Big|_{\,a\to 0}=k(\Box)\equiv 
    \frac14\left(\,\ln\frac4{l^2(-\Box)} 
    -{\bf C}\,\right).                    \label{8.26} 
    \end{eqnarray} 
 
This Wick rotation after moving the second brane to the AdS 
horizon imposes a special choice of vacuum or special boundary 
conditions at the AdS horizon. The Hartle boundary conditions 
corresponding to this type of analytic continuation imply that the 
basis function $u_-(z)$ instead of (\ref{4.6}) is given by the 
Hankel function, 
$u_-(z)=H^{(1)}_2(z\sqrt\Box)=J_2(z\sqrt\Box)+iY_2(z\sqrt\Box)$, 
and thus corresponds to ingoing waves at the horizon 
\cite{GKR,BalGiLa,BalKraLa}. This is equivalent to the replacement 
$Y_1^-,Y_2^-\to1$, $J_1^-,J_2^-\to-i$, in (\ref{8.23}) and, thus, 
justifies the Wick rotation of the above type. Hartle boundary 
conditions and the Euclidean form factor (\ref{8.26}) naturally 
arise when the Lorentzian AdS spacetime is viewed as the analytic 
continuation from the Euclidean AdS (EAdS) via Wick rotation in 
the complex plane of time. Under this continuation the AdS horizon 
is mapped to the inner regular point of EAdS, the coordinate $z$ 
playing a sort of an inverse radius, and the regularity of fields 
near this point is equivalent to the exponential decay of the 
Euclidean basis function $u_-(z)=2iK_2(z\sqrt{-\Box})/\pi\to 0$, 
$z\to\infty$ (remember that $\Box<0$ in Euclidean spacetime). Such 
an analytic continuation models the mechanism of cosmological 
creation via no-boundary \cite{HH} or tunneling \cite{tun} 
prescriptions extended to the braneworld context 
\cite{GarSas,HHR1,HHR2,brane}. In particular, it determines 
(otherwise ambiguous) nonlocal operations in Lorentzian spacetime 
from their uniquely defined Euclidean counterparts.

\section{Conclusions\label{conclusions}} 
 
To summarize, we have constructed the braneworld effective action in the 
two-brane Randall-Sundrum model to quadratic order in curvature perturbations 
and radion fields on both branes. We have obtained the exact nonlocal form 
factors in this two-field action and their low-energy approximation. The zeros 
of the form factors reproduce the spectrum of Kaluza-Klein modes. Thus, this 
description is intrinsically equivalent to the Kaluza-Klein setup.  However, 
it explicitly features two fields rather than the infinite tower of local 
Kaluza-Klein modes. The price one pays for this is that the 
two-field action is essentially nonlocal and its nonlocality is a cumulative 
effect of the Kaluza-Klein modes.  
 
We have also considered the reduced version of the action --- the functional  
of the fields associated 
with only one positive-tension brane. A physical motivation for 
this reduction is the fact that, if this brane with its metric and 
other fields is regarded as the only visible one, then one has to 
trace out in the whole two-brane system the fields on the second 
brane\footnote{This may also lead to decoherence for the 
visible fields, cf.~\cite{Karpacz}.}. 
In the tree-level approximation this procedure is 
equivalent to excluding the invisible fields via their equations 
of motion in terms of the fields on the positive-tension brane.  
In the low-energy 
approximation the result turns out to be very simple --- the action 
is dominated by the Einstein term with the four-dimensional 
gravitational constant explicitly depending on the brane separation. 
This is a manifestation of the well-known localization of the 
graviton zero mode on the brane or recovery of the low-energy 
Einstein theory. 
 
However, for finite interbrane distance the action does not get 
completely localized even in the long-distance approximation --- in 
the conformal sector a certain nonlocality survives.  Nevertheless, 
the latter can be localized in terms of the additional scalar 
field non-minimally interacting with the brane curvature. 
Interestingly, the result coincides with the braneworld action 
constructed in \cite{brane} by a simplified method disregarding 
the metric perturbations on the negative-tension brane. In 
contrast to \cite{brane}, the radion field in (\ref{8.16}) has a 
dipole ghost nature, but on its mass shell the action 
(\ref{8.16}) concides with that of \cite{brane}, given by (\ref{8.19}). 
Moreover, off-shell both actions can be identically 
transformed into one another by the nonlocal reparametrization 
(\ref{8.18}). 
 
Thus, our results justify the conclusions of \cite{brane}.  
These conclusions were used to generate inflation by means of  
the radion field (\ref{8.18}) playing the role of an inflaton. For this 
purpose, the theory was generalized to the case when the tension on 
the visible brane is slightly detuned from the flat brane 
Randall-Sundrum value. Then the action (\ref{8.19}) rewritten in 
the Einstein frame would acquire a nontrivial inflaton potential, 
such that its slow-roll dynamics corresponds to branes diverging 
under the action of a repulsive interbrane force --- a scenario 
qualitatively different from the models of colliding branes  
\cite{DvTyeSh,Ekpyr,brantibr}. However, it suffers 
from an essential drawback. This is the necessity to introduce by hand 
a four-dimensional cosmological constant --- the brane 
tension detuning of the above type.  
 
Quite interestingly, the  
present results suggest the mechanism of interbrane repulsion 
based on the presence of the Weyl-squared term in (\ref{8.16}) and  
(\ref{8.24}). When the brane Universe is filled with the graviton 
radiation this term is nonzero, $C_{\mu\nu\alpha\beta}^2>0$,  
and for small brane separation it forms the interbrane potential 
$-(l^2/2)\kappa_1(a)C_{\mu\nu\alpha\beta}^2$. It has a maximum 
at the point of coinciding branes $a=1$ because the coefficient 
$\kappa_1(a)$ given by (\ref{8.10a}) is strictly positive. The 
repelling force is very small, though, and identically vanishes  
at $a=1$, because of the behaviour of $\kappa_1(a)$ at the brane 
collision point\footnote{Interestingly, the expression (\ref{8.10a}) 
represents  the logarithm $\ln(1/a^2)$ with exactly  
the first two terms of its Taylor series in $(1-a^2)$  
subtracted.}, $\kappa_1(a)\sim(1-a^2)^3/12$. Unfortunately, this 
potential is strictly negative, because $\kappa_1(a)\geq 0$, and,  
therefore, cannot maintain  
inflation (for recent studies of models with a negative cosmological  
constant, see \cite{FFKL}). Rather, it can serve as a basis of brane 
models with $AdS_4$ geometry embedded in $AdS_5$ \cite{AdS_4brane}. 
It can also be useful in the model of ``thick'' branes in the  
Big Crunch/Big Bang transitions \cite{TurToll} of ekpyrotic and 
cyclic cosmologies \cite{Ekpyr,cycle} and provide an alternative  
to the mechanisms of repulsion caused by matter fields on branes  
\cite{Wiseman}. 
 
Thus, the renormalization flow in $a$ (AdS flow) can, in principle,  
be realized in our model at the dynamical level. This flow  
interpolating between  short and long  
interbrane distances is very interesting. It features the 
transition from the phase of the local action (\ref{8.19}) to 
the action (\ref{8.24}) with logarithmic nonlocal form factors in the 
Weyl-squared term. At the initial stage the local logarithm in 
$\kappa_1(a)$, Eq.~(\ref{8.10a}), structurally  
resembles the logarithmic behaviour of the Coleman-Weinberg potential 
of the field (\ref{8.18}), $\varphi\sim\sqrt{3/4\pi G_4}\,a$,  
$\kappa_1(a)=-(1/4)\ln(\varphi^2/m_P^2)+...$, $m_P^2=1/G_4$ 
(although it multiplies instead of the local power of $\varphi$ the 
square of the Weyl tensor). When extrapolated to small $a$ (big  
interbrane distances) it, instead of a naive infinite growth, enters 
another energy domain (\ref{7.4.1}),  
$a^2\sim l^2\Box$, where it gets saturated by the 
characteristic scale of the Weyl tensor (or the energy scale of the 
graviton radiation contained in the model). Therefore, the local  
coefficient $\kappa_1(a)$ gets replaced by the form factor  
$k_1(\Box)$, Eq.~(\ref{8.23}), the dominant logarithmic term of  
$\kappa_1(a)\sim(1/4)\ln(1/a^2)$ going over into the  
logarithm of $k_1(\Box)\sim (1/4)\ln(4/l^2\Box)$. Thus, this  
renormalization AdS flow leads to the delocalization of the 
initial radion condensate $\kappa_1(a)C_{\mu\nu\alpha\beta}^2$ 
to nonlocal (but short-distance) corrections  
$C_{\mu\nu\alpha\beta}\,k_1(\Box)C^{\mu\nu\alpha\beta}$ 
characteristic of the AdS/CFT-correspondence principle in the limit of 
large interbrane distance. 
 
Concrete implications of these phase transitions in cosmology 
still have to be worked out. Here we only make a final remark 
that they can comprise an essential point of departure from the 
scenario of diverging branes of \cite{brane} for large interbrane 
distances (\ref{7.4.1}). The corresponding action (\ref{8.24})  
is different from the one in (\ref{8.19}) (or equivalently (\ref{8.16}))  
that was 
extrapolated in \cite{brane} to late stages of the brane runaway. 
In contrast to (\ref{8.19}), the action (\ref{8.24}) does not have 
any non-minimal curvature coupling of the modulus $a$, which 
together with the brane tension detuning served as a basis for the 
acceleration stage in \cite{brane}. The diverging-branes scenario 
of \cite{brane} also admitted this stage as a sequel to the 
slow-roll inflation, but the inflation and 
acceleration stages overlapped there and, thus, caused 
unsurmountable difficulties for reheating \cite{FFKL}. Thus, it 
would be interesting to observe the effect of the nonlocal  
short-distance corrections (replacing the non-minimal curvature  
coupling of (\ref{8.19})) on the late time behavior in the scenario of 
diverging branes. Since these corrections are dominated by 
curvature-squared terms, their effect can be equivalent to the 
$R^2$-inflation model \cite{R^2} (see also \cite{Muk} for the same 
conjecture on the realization of the Starobinsky model in 
braneworld scenarios). We hope to deal with these issues in a 
future publication. 
 
\section*{Acknowledgements} 
 
One of the authors (A.O.B.) benefitted from helpful 
discussions with V.~Rubakov, R.~Metsaev and S.Solodukhin. The authors 
are indebted to D.~Nesterov for checking some of the calculations in 
this paper. 
A.O.B.\ and A.Yu.K.\ are grateful for the 
hospitality of the Theoretical Physics Institute, University of Cologne, where  
a major part of this work has been done due to the support of the DFG grant  
436 RUS 113/333/0-2.  
A.Yu.K.\ is also grateful to CARIPLO Science Foundation.  
The work of A.O.B.\ was also supported by the Russian  
Foundation for Basic Research under the grant No 02-02-17054 and the  
grant for Leading scientific schools No 00-15-96566, while  
A.Yu.K.\ was supported by the RFBR grants No 02-02-16817 and No 00-15-96699. 
 A.R.\ is supported by the DFG Graduiertenkolleg  
``Nonlinear Differential Equations''.

\end{document}